# The FAST Framework: Developing a Data-Efficient Machine Learning Potential to Decode Superionic Transition-Induced Thermophysical and Kinetic Anomalies in $UO_2$ under Extreme Conditions

Fengnian Zhuang[1] Gaosheng Yan[1]* Hong Chen[2] Yi Zhang[1,2] Wenshan Yu[1]*

Minglong Xu[1] Shengping Shen[1]

[1]State Key Laboratory for Strength and Vibration of Mechanical Structures, Shaanxi Engineering Laboratory for Vibration Control of Aerospace Structures, School of Aerospace, Xi'an Jiaotong University, Xi'an 710049, China

[2]National Key Laboratory of Strength and Structural Integrity, Aircraft Strength Research Institute of China, Xi'an 710065, China

*corresponding to: ygs@xjtu.edu.cn (G.S. Yan) wenshan@xjtu.edu.cn (W.S. Yu)

**Abstract:** Uranium dioxide ($UO_2$) serves as the predominant nuclear fuel globally. Despite its widespread application, evaluating its mechanical, thermophysical, and species transport behaviors under extreme accident scenarios remains a formidable challenge for conventional experimental and computational methods. To address this, we develop a versatile machine learning interatomic potential (MLIP) for $UO_2$ by proposing an efficient training strategy, termed “FAST” (**F**ine-tuning via **A**ctive-learning and **S**uperionic-**T**argeting) framework. Our "FAST" framework integrates superionic transition-targeted sampling with active learning-enhanced exploration to efficiently construct a highly compact dataset comprising only 500 configurations for fine-tuning a foundation model. By rigorously accounting for the strong correlation of uranium 5*f* electrons and antiferromagnetic (AFM) ground state during DFT labeling,

`

we train a robust DFT-level neuroevolution potential (NEP) for $UO_2$. We demonstrate that this NEP exhibits superior predictive capability for various physical properties, encompassing mechanical, defect, thermophysical, and ionic diffusion over an extensive temperature range. Moreover, this NEP accurately captures the anomalous thermophysical and kinetic behaviors triggered by superionic transition. Specifically, it reproduces both "λ-peak" in linear thermal expansion coefficient (LTEC) and "non-Arrhenius" anionic diffusion. Crucially, NEP-based simulations elucidate the microscopic origins underlying these anomalies: the pre-melting of oxygen sublattice and resultant kinetic decoupling between U and O ions. This work not only provides a high-fidelity NEP for evaluating the thermo-mechanical-diffusion properties of $UO_2$ under extreme conditions, but also proposes a highly efficient "FAST" framework. This transferable framework may pave the way for developing robust MLIPs for other materials exhibiting complex phase transitions.

## 1. Introduction:

Uranium dioxide ($UO_2$) is the primary commercial nuclear fuel utilized worldwide. During in-reactor service, it is subjected to extreme temperatures, severe irradiation, and significant mechanical stress, inducing structural degradation that limits its operational reliability and lifespan. Therefore, a comprehensive understanding of the mechanical and thermophysical property variations of $UO_2$ under varying conditions is essential for assessing its operational performance and guiding the rational microstructural design to extend its service life. Extensive experimental efforts have been devoted to examine the mechanical [1, 2], species transport [3, 4], and thermophysical properties [5, 6] of $UO_2$. However, these studies predominantly focus on the conventional operating temperatures (600–1500K) [7], while precise investigations under extreme accident conditions (approaching the melting point) remain relatively scarce. The lack of high-temperature data primarily stems from the prohibitive costs and immense technical challenges of sustaining stable ultra-high temperatures over extended periods, as well as accurately measuring the related physical properties of such an inherently radioactive and hazardous material. To bridge this gap, theoretical simulation methods including the density functional theory (DFT) and molecular dynamics (MD) are utilized. Despite the DFT can accurately simulate certain static properties of $UO_2$ [7, 8], its computational expense is severely exacerbated by the strong correlation of uranium 5*f* electrons [9]. In addition, converging to the correct magnetic ground state of $UO_2$ is another notoriously intractable issue, as the standard DFT calculations frequently become erroneously trapped in metastable magnetic states [7, 10]. Accordingly, these electronic complexities preclude the extensive applications of DFT to perform large-scale and long-timescale dynamical simulations of $UO_2$. Alternatively, numerous empirical interatomic potentials (EIPs) [11-15] have been developed and applied to investigate various dynamics properties of $UO_2$, such as thermodynamics [16, 17], diffusional creep [18, 19], stress-induced phase transition and fracture [17, 20]. However, the predictive fidelity of MD fundamentally relies on an accurate description of the potential energy surface (PES). Constrained by their rigid analytical forms and a scarcity of high-temperature experimental data, traditional EIPs for $UO_2$ are parameterized primarily against low-temperature properties. Consequently, their capability to reliably extrapolate to the complex high-energy PES near the melting point is inherently restricted, severely limiting their transferability and accuracy under extreme conditions [21].

In recent years, machine learning interatomic potentials (MLIPs) have emerged as a promising paradigm, achieving near-DFT accuracy at classical EIPs-level computational costs. To date, various MLIPs have been developed for $UO_2$, demonstrating first principles fidelity for specific properties. For instance, neural network potentials (NNPs) and spectral neighbor analysis potentials (SNAPs) have been successfully developed to accurately predict diverse thermophysical properties of $UO_2$ [22, 23]. Building upon these existing MLIP frameworks, the integration of active learning has further enhanced the training efficiency and bolstered the predictive accuracy of $UO_2$ potentials. For instance, Zhong *et al*. have constructed an active learning-enhanced neuroevolution potential that accurately evaluates the lattice thermal conductivity of $UO_2$ within the low-to-intermediate temperature regime (300–1500 K) [24]. Moreover, focusing on assessing the quality of training dataset construction, a recent study investigates the necessity of incorporating Hubbard corrections (DFT+U) during the training dataset labelling. By systematically benchmarking MLIPs trained on datasets derived from standard DFT and DFT+U calculations against experimental measurements, the researchers demonstrate the necessity of introducing the Hubbard corrections (DFT+U) for accurate reference labeling. Actually, unlike the standard DFT calculations, incorporating Hubbard corrections with appropriate values of Hubbard parameter $U$ and exchange parameter $J$ can effectively capture the strong correlation of uranium 5*f* electrons in $UO_2$ [25]. Such ensures a physically rigorous and high-quality training dataset, thereby enabling the trained MLIPs to accurately reproduce the thermophysical and mechanical properties of $UO_2$ [10]. Nevertheless, when studying $UO_2$ under extreme conditions, developing an MLIP capable of accurately and robustly capturing the anomalous superionic transition (ST) would introduce extensive complexities that transcend the routine training dataset generation.

The ST, also known as the Bredig, lambda or Faraday transition [26, 27], is a second-order phase transition widely observed in $UO_2$ near 0.85 $T_m$ ( $T_m$ : melting point) through both experimental measurements [28] and theoretical simulations [29, 30]. In-depth investigations into the ST hold dual significances. From a macroscopic engineering perspective, the ST is activated by the synergistic interplay of thermodynamic anomalies (e.g., "λ-peak" of linear thermal expansion coefficient), intensified species diffusion, and localized pre-melting [10, 19]. Moreover, as demonstrated in our previous studies, the ST is strongly linked with the transitions in the high-temperature diffusional creep behavior of $UO_2$ [18, 19]. Thus, understandings

`

of the ST are of paramount importance for analyzing coupled thermo-mechanical-diffusional responses and predicting fuel service life under severe accident conditions. From a theoretical simulation perspective, accurately reproducing the ST requires exceptionally stringent demands on the fidelity of the PES in far-from-equilibrium regimes of phase space. In fact, successfully reproducing the ST stands as the "Gold Standard" for validating an $UO_2$ MLIP's predictive accuracy under extreme conditions. To meet this rigorous standard, active learning (AL)-enhanced iterations naturally emerge as the undisputed choice to efficiently explore the highly disordered atomic configurations in the extreme superionic states. Conventionally, developing a robust MLIP requires multiple AL-enhanced exploration iterations to construct massive training datasets, typically comprising at least several thousand atomic configurations, followed by hundreds of thousands (100,000) to as many as one million training steps (1,000,000) [10, 23, 24]. For training a $UO_2$ MLIP operating at elevated temperatures, executing this standard workflow on highly disordered atomic configurations imposes prohibitive computational burdens during both DFT labeling and MLIP fitting. To overcome this hurdle, a novel and increasingly adopted strategy has emerged. This approach utilizes advanced AL-enhanced iterations to efficiently explore highly diverse atomic configurations across the phase space, thereby selecting and consolidating them into a highly compact dataset comprising remarkably few representative structures [31]. Subsequently, a pre-trained universal foundational model is fine-tuned based on this compact dataset for merely 10,000 to 20,000 steps. Compared to conventional direct training, this strategy provides a significantly more economical and efficient pathway to generating robust MLIPs for targeted materials [32, 33].

In this study, we develop a high-fidelity neuroevolution potential (NEP) [34] for $UO_2$ via our proposed "FAST" (**F**ine-tuning via **A**ctive-learning and **S**uperionic-**T**argeting) training framework. Within "FAST", we first construct a highly compact dataset of merely 500 representative atomic configurations, which is generated through ST-targeted sampling and AL-enhanced exploration processes. Subsequently, these configurations undergo high-accuracy DFT labeling, where the strong correlation of uranium 5*f* electrons and the antiferromagnetic (AFM) ground state are rigorously

`

treated. These processes establish a high-fidelity dataset, which is then used to fine-tune the universal NEP89 foundational model, ultimately yielding a robust DFT-level NEP for $UO_2$. The predictive accuracy of this NEP is systematically benchmarked against experimental and theoretical calculation data (e.g., DFT, MD) across diverse mechanical, defect, thermophysical, and ionic diffusion properties. Furthermore, the microstructural origins activating the ST are elucidated through the pre-melting of the oxygen sublattice, as evidenced by its pair distribution function (PDF) variations. Further comparative analysis of the ionic diffusion reveals that the ST-induced structural evolution drives a profound kinetic decoupling between the uranium and oxygen ions, thereby resulting in the onset of anomalous "non-Arrhenius" anionic diffusion. More importantly, the ST phenomena are ubiquitous across diverse material systems, including fluorite and anti-fluorite ceramics [35, 36], solid-state electrolytes [37-39], thermoelectrics [40, 41], and planetary interiors [42, 43]. Therefore, beyond the specific $UO_2$ NEP, the proposed "FAST" framework may provide a highly transferable strategy for efficiently developing potentials for materials undergoing similar complex phase transitions under extreme conditions.

## 2. "FAST" training framework

### 2.1 Construction of training dataset

#### 2.1.1 Baseline dataset preparation

To prepare a preliminary model via fine-tuning the universal NEP89 foundation model, a baseline dataset is first constructed using the NepTrainKit software [44]. Firstly, a fluorite-type 2×2×2 $UO_2$ supercell comprising 96 atoms (32 uranium and 64 oxygen atoms) is established. Random atomic displacement perturbations of 0.3 Å, coupled with uniform lattice scaling factors ranging from 0.96 to 1.04, are applied to this supercell. Subsequently, lattice deformations encompassing the uniaxial, biaxial, and triaxial strains within ±5% are imposed. In addition, shear matrix strains (-3%~3%) and shear angle strains (-2°~2°) are applied to enhance the capability of the preliminary model in describing the simulation box distortions. To further equip the preliminary model to accurately capture far-from-equilibrium defect evolution and defect-mediated ion diffusion, we extend the baseline dataset with defect-containing configurations.

Specifically, Schottky defects (SDs) comprising 1 to 3 pairs in both bound and isolated states are randomly introduced into the 2×2×2 supercells. These SDs are selected to preserve the charge neutrality within the supercells, which is achieved by removing uranium and oxygen atoms in a stoichiometric 1:2 ratio. Supercells containing surface slabs are also incorporated into the baseline dataset, featuring randomly sampled Miller indices (*h*, *k*, *l*) ranging from 0 to 3, thicknesses of 3 to 6 atomic layers, and a fixed 10 Å vacuum gap. To efficiently map the true physical PES of $UO_2$ and substantially enhance the trained potential's robustness during dynamics simulations, extensive MD simulations are performed by using the LAMMPS [45] and GPUMD software [46]. These MD simulations utilize two established $UO_2$ interatomic potentials (a Neural network potential [23] and CRG empirical potential [14]), alongside the NEP89 pre-trained foundation model covering 89 elements [33]. All simulations are conducted on 2×2×2 $UO_2$ supercells under both the NVT and NPT ensembles, employing the Nosé-Hoover thermostat and Nosé-Hoover barostat. To ensure a comprehensive exploration of the phase space, these MD simulations are performed to fully span temperatures from 300 to 5000 K and pressures from -5 to 5 GPa. The timestep is set to 1 fs, yielding a total simulation duration of 500 ns. A specific number of atomic snapshots are uniformly sampled from the MD simulation trajectories. Subsequently, to ensure dataset plausibility, unphysical structures with unrealistic interatomic bond lengths are identified and filtered out using the NepTrainKit software. Ultimately, these chosen dynamic snapshots are pooled with the previously generated static configurations. A farthest point sampling (FPS) method is then applied to further select the most representative structures, thereby maximizing the structural diversity and minimizing dataset redundancy [44]. In total, the baseline dataset comprises 240 configurations: 20 perturbed, 37 sheared, 22 defective, 37 surface and 124 MD-sampled structures.

### 2.1.2 Superionic transition (ST)-targeted sampling

To further enhance the NEP's ability for high-temperature applications, an ST-targeted sampling strategy is introduced to augment the baseline dataset. Physically, the ST in $UO_2$ is characterized by the pre-melting of oxygen sublattice while the uranium sublattice remains crystalline. At the onset of ST, O-ions transition into a highly mobile, liquid-like state, whereas U-ions oscillate around their equilibrium sites, resulting in the diffusivity of O-ions remarkably higher than that of U-ions. To empower the NEP to reproduce the accurate PES features governing the ST, it is crucial to assign a higher sampling weight to configurations near the onset of ST. Recognized for its high accuracy in reproducing the ST-related properties of $UO_2$, CRG potential is adopted to supplement the MD sampling within the ST regime. To comprehensively capture the solid-superionic-liquid transition, continuous temperature-ramped MD simulations are conducted. Considering the ST as a gradual second-order transition [29] with an onset near 2600 K for the CRG potential [17, 47], the 2×2×2 supercells are continuously heated from 2200 to 2800 K over 50 ns under a zero-pressure NPT ensemble. Atomic snapshots are uniformly extracted from the resulting MD trajectories. Subsequently, the FPS method is applied to select the most structurally diverse configurations from this ST-sampled pool relative to the baseline dataset. Ultimately, 60 configurations extracted from this ST-enhanced sampling scheme are integrated into the baseline dataset to form an augmented baseline dataset of 300 configurations. To validate the effectiveness of this ST-targeted strategy in expanding phase-space coverage, the high-dimensional structural descriptors of the training dataset are visualized using dimensionality reduction techniques. Specifically, the Principal Component Analysis (PCA) and Uniform Manifold Approximation and Projection (UMAP) projections are detailed in Section S1 of the Supplementary Material. Utilizing this augmented baseline dataset, the initial NEP model (denoted as NEP0) is trained by fine-tuning the NEP89 foundation model for 20,000 steps. The Introduction of NEP potential and detailed hyperparameters for this fine-tuning process are provided in Section S2~S3 of the Supplementary Material.

**2.1.3 Active-learning (AL)-enhanced exploration**

To systematically empower the initial NEP0 with superior generalization and phase-space exploratory capability, a progressive two-stage active learning (AL) framework is introduced. Building upon the augmented baseline dataset, AL-enhanced Stage 1 focuses on expanding the training set into unexplored phase spaces. The trained NEP0 is utilized to conduct extensive MD sampling on 2×2×2 $UO_2$ supercells across a broad thermodynamic regime (NVT/NPT ensembles, 300~5000 K, -5~5 GPa). From these simulations, atomic snapshots are uniformly extracted to form a vast candidate pool. Subsequently, the FPS algorithm implemented in the GPUMDkit package [48], is applied to minimize structural redundancy by filtering the newly sampled pool against the augmented baseline dataset. This FPS selection successfully extracts the 80 most structurally diverse configurations, denoted as the FPS dataset. Based on this subset, the Maxvol algorithm is applied to construct a representative active set. The remaining candidate structures are then evaluated using the D-optimality criterion. Specifically, configurations with a $\gamma$ extrapolation grade falling between 1 and 2 ( $\gamma \in [1,2]$ ) are identified as reliable extrapolation candidates [49]. Exactly 70 of these D-optimality-selected structures (D-optimality dataset) are merged with the FPS dataset to form the Stage 1 dataset, comprising a total of 150 new structures. Consequently, the total size of the training dataset is expanded from 300 to 450 configurations.

Following the efficient phase space exploration achieved in Stage 1, a second AL-enhanced exploration Stage 2 is conducted to extrapolate configurations that are difficult to capture through purely FPS and D-optimality criterion. To accomplish this high-uncertainty sampling, the Query-by-Committee (QBC) method, also known as the ensemble method is employed [50]. Initially, the 450-configuration dataset is randomly split into an 80% training set and a 20% test set. A committee of five NEP models (NEP1–NEP5) is established by fine-tuning the NEP89 foundation model for 20,000 steps with distinct random seeds. Subsequently, MD simulations are conducted using NEP1 over a wide temperature range (300~3000K) and pressure range (-5~5GPa) under the NPT ensemble for a total of 10ns. Concurrently, on-the-fly uncertainty evaluations are performed by the NEP1~NEP5 committee at uniform 100ps intervals during these MD simulations. The prediction uncertainty is quantified by the maximum standard deviation of atomic forces among the committee for a given configuration ( $\sigma_f$ ) (The definition of the $\sigma_f$ criterion is detailed in Section S4 of the Supplementary materials).

`

Configurations exhibiting severe disagreement among the NEP1~NEP5 committee, which is defined by $\sigma_f$ exceeding a threshold of $\delta$=0.1 eV/Å ($\sigma_f > \delta$), are selected for inclusion in the training dataset. Eventually, 40 configurations with the highest $\sigma_f$ are supplemented, yielding a final training dataset containing 500 configurations for developing the final $UO_2$ NEP. Note that to guarantee the physical validity of the configurations in the training dataset, all supplemented configurations from each process are filtered using the NepTrainKit software to remove unphysical structures prior to high-accuracy DFT labeling. The overall phase-space coverage and distributions of configurations from distinct dataset origins are presented in Section S2 of the Supplementary materials.

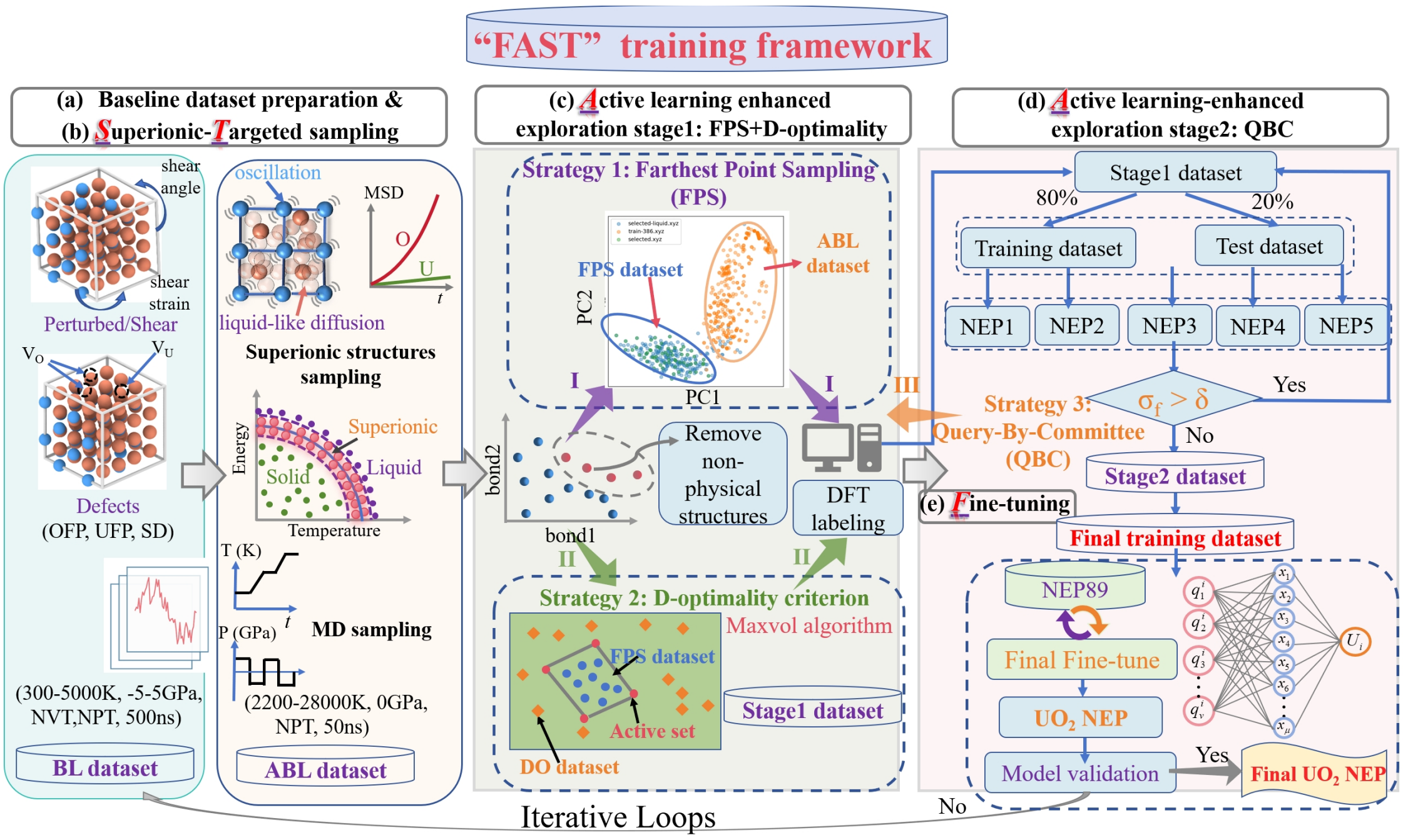

**Fig. 1**. Schematic of the "FAST" (**F**ine-tuning via **A**ctive-learning and **S**uperionic-**T**argeting) training framework for the $UO_2$ NEP. The pipeline integrates (a) baseline dataset preparation, (b) superionic-targeted sampling, (c) active learning-enhanced exploration stage 1 (FPS + D-optimality), (d) active learning-enhanced exploration stage 2 (QBC), and (e) fine-tuning of the final $UO_2$ NEP from the NEP89 foundation model. BL, ABL, and DO denote the baseline, augmented baseline, and D-optimality datasets, respectively. Pathways I, II, and III indicate that configurations selected by the three AL-enhanced exploration strategies must be filtered to remove non-physical

`

structures prior to high-accuracy DFT labeling.

The comprehensive “FAST” training framework is summarized in Fig. 1. Notably, the acronym “FAST” is originated from the reverse sequence of this workflow, which consists of five sequential steps: (a) baseline dataset preparation, (b) superionic-targeted sampling, (c, d) active learning-enhanced exploration stages 1 and 2, and (e) fine-tuning of the NEP89 foundation model.

## 2.2 NEP Training process

### 2.2.1 DFT calculation settings

To provide ground-truth labels for the NEP training, high-accuracy self-consistent field (SCF) calculations are performed on all sampled configurations to evaluate their energies, atomic forces, and virial tensors. The training dataset consists of totally 500 representative $2\times2\times2$ supercells (96 atoms each). All SCF calculations are conducted using the Vienna ab initio Simulation Package (VASP) [51]. To accurately capture the experimentally observed antiferromagnetic (AFM) ground state of $UO_2$, spin polarization is considered in the SCF calculations. Specifically, a collinear 1k AFM order (without spin–orbit coupling (SOC)) is impose [52], wherein the initial magnetic moments for the uranium atoms are set to +2 and -2 $\mu_B$, while the oxygen atoms are kept at 0 $\mu_B$ [10, 23]. The electron-ion interactions and exchange-correlation effects are described using the projector augmented-wave (PAW) method [53] and the generalized gradient approximation (GGA) parameterized by the Perdew–Burke–Ernzerhof (PBE) functional [54], respectively. A plane-wave cutoff energy of 550 eV is adopted, alongside Gaussian smearing with a broadening width of $\sigma$=0.05 eV. The Brillouin zone is sampled utilizing a Γ-centered Monkhorst–Pack (MP) *K*-point mesh [55]. Specifically, a 2×2×2 *K*-point mesh grid is employed for all configurations. Furthermore, to accurately account for the strongly correlated behavior of the 5*f* electrons in $UO_2$, a Hubbard parameter correction (DFT+U) is incorporated into the standard exchange-correlation functional. Notably, although DFT+U successfully reproduces the experimentally observed AFM ground state and Mott-insulating nature of $UO_2$ [10, 56], these calculations are highly prone to converge to metastable states [7]. To address this, methods such as $U$-ramping and occupation matrix control (OMC) have been proposed to ensure robust convergence to the true global minimum energy state [57, 58]. Considering the extensive number of SCF calculations required for labelling the dataset, the highly efficient $U$-ramping approach is adopted. Specifically, the Hubbard parameter $U$ is gradually increased from 0.51 to 4.51 eV in an increment of 1.0 eV, while the exchange parameter $J$ is held constant at 0.51 eV. Consequently, the effective Hubbard parameter ($U_{eff} = U - J$) is ramped from 0 to a final value of 4.0 eV, aligning with the widely accepted values in the literature for $UO_2$ [10, 23, 56]. The energy convergence threshold for all SCF calculations is set to $10^{-6}$ eV.

**2.2.2 Final fine-tuning based on the NEP89 foundation model**

Utilizing the same set of hyperparameters as those used for training the initial NEP0, the final $UO_2$ NEP is trained on the complete dataset of 500 configurations by fine-tuning the NEP89 foundation model for 20,000 steps. The cutoff radii for the radial and angular descriptors are set to 6 Å and 5 Å, respectively. To enhance the NEP model's descriptive capability for highly disordered configurations at extreme temperatures, and to accurately capture atomic interactions during high-energy irradiation collisions, a Ziegler-Biersack-Littmark (ZBL) screened nuclear repulsion potential is incorporated [59, 60]. The detailed integration scheme of the NEP and ZBL potentials is provided in Section S5 of the Supplementary materials. The inner and outer cutoffs for the ZBL potential are defined as 1 Å and 2 Å, respectively. During the fine-tuning process, the loss weights for energies, atomic forces, and virial tensors are all set to 1. The neural network architecture comprises a single hidden layer with 80 neurons, and both L1 and L2 regularizations are applied to promote generalization. Clearly, as shown in Fig. 2(a), the loss functions of energies, atomic forces, and virial tensors decrease monotonically and converge to stable plateaus during the fine-tuning process. And the robustness of the NEP is evidenced by the convergence of L1 and L2 regularization terms. Furthermore, our NEP predictions aligns well with the DFT+U reference values, achieving root-mean-square errors (RMSEs) of 9.09 meV/atom and 361.68 meV/Å for energies, and forces, respectively (Figs. 2(b)–(d)). To facilitate the subsequent evaluation of mechanical properties, the virial tensor error is converted and reported as a stress RMSE, yielding a remarkably low value of 0.46 GPa (Fig. 2(d)). Notably, the NEP-predicted stresses span a broad range from −10 to 40 GPa, indicating both the potential's capability to accurately describe diverse configurational environments and its robust performance under relatively extreme conditions. Further details regarding the fitting accuracy of the NEP, as evidenced by density-based parity plots and density of atomistic states (DOAS) profiles, are provided in Section S6 of the Supplementary Material.

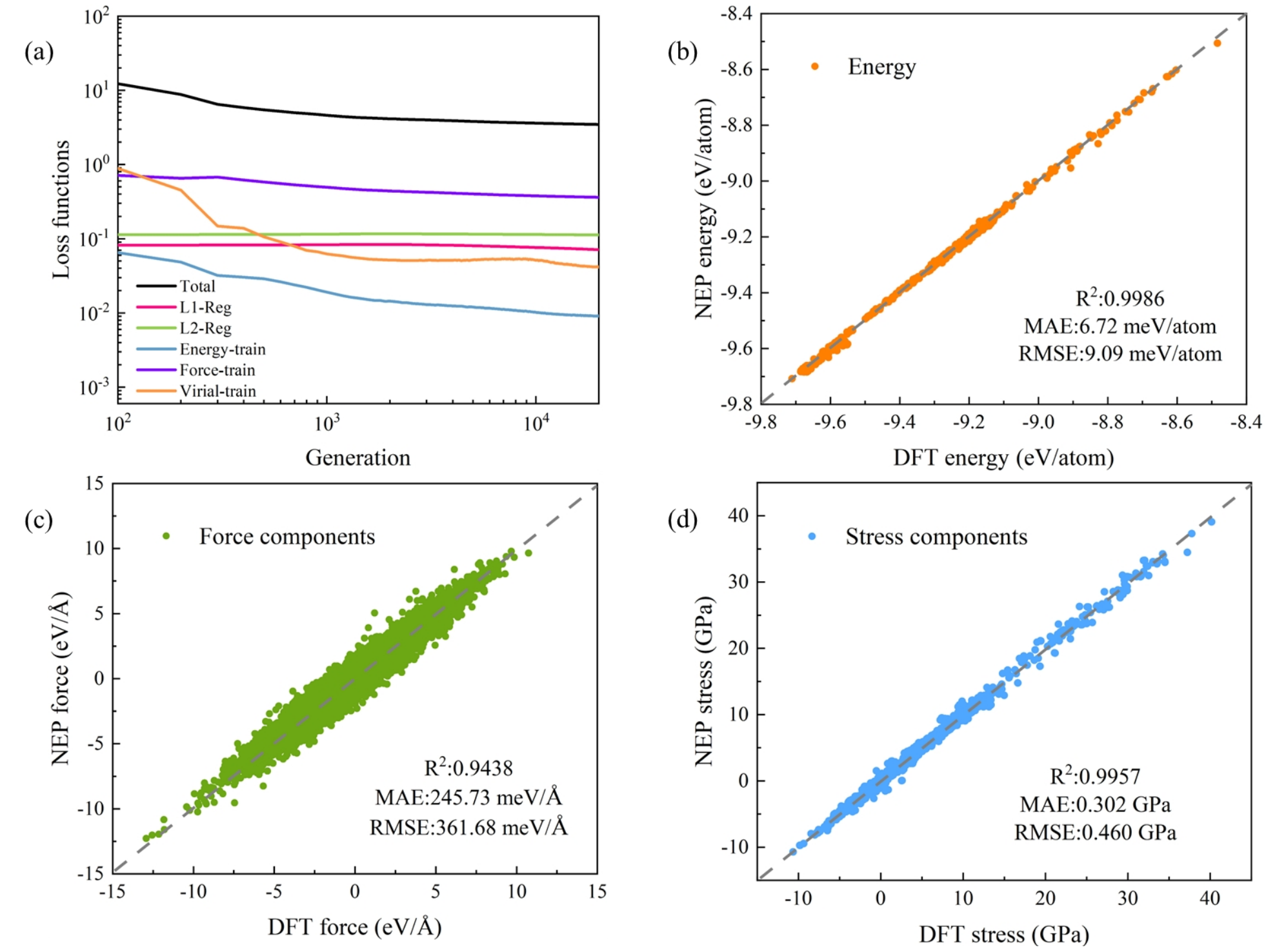

**Fig. 2** Fine-tuning results of the $UO_2$ NEP. (a) Convergence of the loss curves for energies, forces, and stresses, along with the L1 and L2 regularization terms during training. (b)~(d) Parity plots comparing the NEP predictions versus the reference DFT calculations for (b) energies, (c) forces, and (d) stresses. Values of the coefficient of determination ($R^2$), mean-absolute-errors (MAE) and root-mean-square errors (RMSE) are marked in the subplots.

## 3. Results:

All MD simulations are conducted using the GPUMD software [46] with an integration time step of 1 fs. Periodic boundary conditions (PBCs) are imposed to all three mutually orthogonal directions of models. Notably, GPUMD is selected as the MD engine primarily due to its native compatibility with the NEP model, enabling exceptional computational efficiency through the GPU acceleration. Unless otherwise emphasized, the simulated system consists of a conventional 7 × 7 × 7 fluorite $UO_2$ supercell containing 4116 atoms. Static energy minimization is performed via the fast inertial relaxation engine (FIRE) algorithm. The Bussi-Donadio-Parrinello (BDP) thermostat [61] and stochastic cell rescaling (SCR) barostat [62] are adopted for the NVT and NPT ensembles, respectively.

**3.1 Elastic and acoustic properties**

The elastic and acoustic properties of single-crystalline $UO_2$ are evaluated via static calculations at 0 K using the Atomic Simulation Environment (ASE) [63] and calorine [64] packages. Firstly, the equilibrium lattice constant ( $a$ ) is determined by fully optimizing the bulk cell until the maximum atomic forces fall below $10^{-4}$ eV/Å. Based on this optimized configuration, the elastic stiffness tensor ( $C_{ij}$ ) is calculated, which is used to obtain Zener's anisotropy factor (Eq.(1)). Furthermore, the long-wavelength acoustic phonon group velocities near the Brillouin zone center ( $\Gamma$-point) are extracted through the finite-displacement method implemented in the calorine package to cross-validate the directional sound speed. Using Eqs. (2)~(5), the isotropic bulk modulus ( $B$ ), and shear modulus ( $G$ ) are derived based on the Voigt–Reuss–Hill (VRH) approximation. Subsequently, Young's modulus ( $E$ ) and Poisson's ratio ( $\upsilon$ ) are obtained via the Eq. (6) and Eq. (7), respectively.

$$Z = \frac{2C_{44}}{C_{11} - C_{12}} \tag{1}$$

$$B = (C_{11} + 2C_{12}) / 3 \tag{2}$$

$$G_{Voigt} = \left(C_{11} - C_{12} + 3C_{44}\right) / 5 \tag{3}$$

$$G_{Reuss} = \frac{5\left(C_{11} - C_{12}\right)C_{44}}{4C_{44} + 3\left(C_{11} - C_{12}\right)} \tag{4}$$

`

$$G = \left(G_{\mathrm{Reuss}} + G_{Voigt}\right)/2 \tag{5}$$

$$E = \frac{9BG}{3B+G} \tag{6}$$

$$\upsilon = \frac{1}{2}(1-\frac{E}{3B}) \tag{7}$$

For the acoustic properties, the longitudinal ( $v_l$ ) and shear ( $v_s$ ) sound velocities are calculated using Eqs. (8)~(9),

$$v_l = \sqrt{(B+\frac{4}{3}G)/\rho} \tag{8}$$

$$v_s = \sqrt{G/\rho} \tag{9}$$

where $\rho$ is the mass density of $UO_2$ (10,545 kg/m$^3$) [65]. The $v_l$ and $v_s$ calculated by the NEP are 5232.89 m/s and 2605.14 m/s, respectively. The Debye temperature ( $\Theta_D$ ) is then calculated by,

$$\Theta_D = \frac{h}{k_B}\left[\frac{3}{4\pi}(\frac{N_A\rho}{M})\right]^{\frac{1}{3}}\left[\frac{1}{3}(\frac{2}{v_s^{\ 3}}+\frac{1}{v_l^{\ 3}})\right]^{-\frac{1}{3}} \tag{10}$$

where $h$, $k_B$, $N_A$, and $M$ denote the Planck's constant, Boltzmann's constant, Avogadro's number, and the average molecular weight, respectively. The Debye temperature ( $\Theta_D$ ) of $UO_2$ predicted by our NEP is 360 K. As summarized in Table 1, all NEP-calculated elastic and acoustic properties of $UO_2$ are in close agreement with both experimental measurements and theoretical calculations utilizing PBE+U, other MLIPs (MLIP-DFT+U [10],SNAP, HDNNP [22, 66]), and EIPs (CRG, SMTB-Q). Note that, the equilibrium lattice constant ( $a$ ) predicted by our NEP model is 5.54 Å, which is slightly larger than those obtained from EIPs (~5.45 Å) and experiments (~5.47 Å). In fact, this slight overestimation is a inherent characteristic stemming from the PBE functional and Hubbard $U$ correction adopted to construct the NEP's training dataset [10, 67]. As expected, our predicted equilibrium lattice constant aligns closely with the values from direct PBE+U calculations and other MLIPs trained on comparable reference data (e.g., MLIP-DFT+U and SNAP).

`

**Table.1** Comparison of the elastic and acoustic properties of $UO_2$ calculated using our NEP model with those derived from other MLIPs, EIPs, first-principle calculations, and experimental measurements.

| Property | NEP (This work) | MLIP-DFT [10] | MLIP-DFT+U [10] | SNAP [22, 66] | HDNNP [22] | CRG [14] | SMTB-Q [15] | PBE+U [68] | EXP 1.[69, 70]. | EXP 2.[71] |
|---|---|---|---|---|---|---|---|---|---|---|
| $a$ (Å) | 5.54 | 5.45 | 5.51 | 5.53 | 5.48 | 5.45 | 5.45 | 5.55 | 5.47 | 5.47 |
| $C_{11}$ (GPa) | 347.4 | 389.47 | 350.82 | 348.9 | 373 | 406 | 398.7 | 343.1 | 396 | 389.3 |
| $C_{12}$ (GPa) | 116.3 | 121.21 | 103.5 | 110.6 | 121 | 124 | 117.6 | 121.3 | 121 | 118.7 |
| $C_{44}$ (GPa) | 51.4 | 77.98 | 47.89 | 52.2 | 64 | 66 | 63 | 62.7 | 64 | 59.7 |
| B (GPa) | 193.3 | 210.63 | 185.94 | 190 | 205 | 218 | 211.3 | 195.23 | 213 | 209 |
| G (GPa) | 71.57 | 97.05 | 70.82 | 73.16 | 84 | 90 | 87.43 | 78.94 | 87 | 83 |
| E (GPa) | 191.1 | 252.39 | 188.52 | 194.51 | 222 | 237 | 230.50 | 208.69 | 230 | 220 |
| Z | 0.445 | 0.582 | 0.387 | 0.438 | 0.50 | 0.47 | 0.448 | 0.565 | 0.46 | 0.441 |
| υ | 0.335 | 0.3 | 0.331 | 0.329 | 0.32 | 0.319 | 0.318 | 0.322 | 0.32 | 0.324 |
| $\Theta_D$ | 360 | 416.89 | 357.53 | 363.31 | 389.35 | 402.22 | 396.59 | 377.02 | 396.59 | 385 |

Prior to investigating the mechanical response of $UO_2$ at varying temperatures, it is imperative to benchmark the temperature-dependent variations of its elastic constants. Accordingly, the variations of the temperature-dependent elastic stiffness tensor ($C_{ij}$) for single-crystalline $UO_2$ is determined utilizing the strain fluctuation approach [31, 72]. By analyzing the fluctuations of the unit cell vectors during MD simulations under the isothermal-isobaric (NPT) ensemble, the finite-temperature elastic constants are derived as,

$$C_{ijkl}^{-1} = \frac{V}{k_B T} \left\langle \varepsilon_{ij} \varepsilon_{kl} \right\rangle - \left\langle \varepsilon_{ij} \right\rangle \left\langle \varepsilon_{kl} \right\rangle \tag{11}$$

where $\varepsilon$ represents the strain tensor of unit cell, $V$ is the volume of unit cell, $T$ is the absolute temperature and, $k_B$ is the Boltzmann's constant. $\langle\ \rangle$ denotes the ensemble average, which is practically evaluated as the time average over the MD trajectory. Notably, to ensure the correct sampling of volume and strain fluctuations at target temperatures, the robust first-order stochastic cell rescaling (SCR) barostat is adopted, which introduces an appropriate stochastic noise term [62].

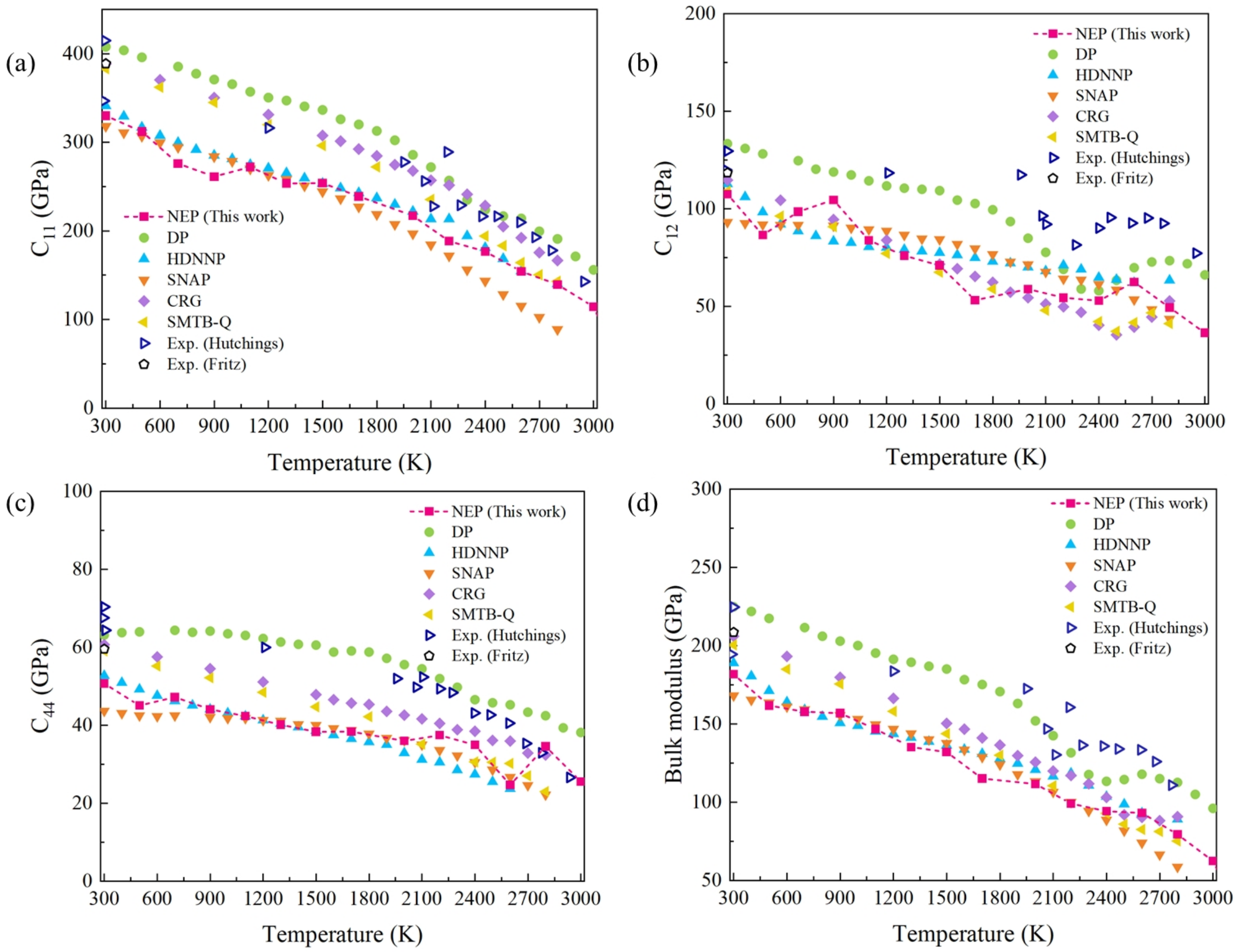

**Fig. 3** Temperature dependence of the elastic constants (a) $C_{11}$, (b) $C_{12}$, (c) $C_{44}$, and (d) bulk modulus of single-crystalline $UO_2$ at the temperature of 300-3000 K. The NEP model predictions (pink dashed lines) are compared with those from other MLIPs (DP [23], HDNNP, SNAP [22]), EIPs (CRG [14], SMTB-Q [15]), and experimental measurements (Hutchings [73], Fritz [71]).

As evidenced in Fig. 3, the elastic constants ($C_{11}$, $C_{12}$, $C_{44}$) and bulk modulus of single-crystalline $UO_2$ undergo significant thermal softening from 300 to 3000 K, which is driven by the enhanced lattice anharmonicity as temperatures increase. Notably, the NEP model predictions (pink dashed lines) align remarkably well with those derived from HDNNP and SNAP MLIPs, while exhibiting modest deviations (10~15%) from the CRG and SMTB-Q EIPs. Moreover, although a somewhat larger offset is observed compared to the experimental measurements by Hutchings and Fritz *et al.*, our NEP model faithfully reproduces the overall decreasing trend of the elastic constants and bulk modulus with increasing temperature.

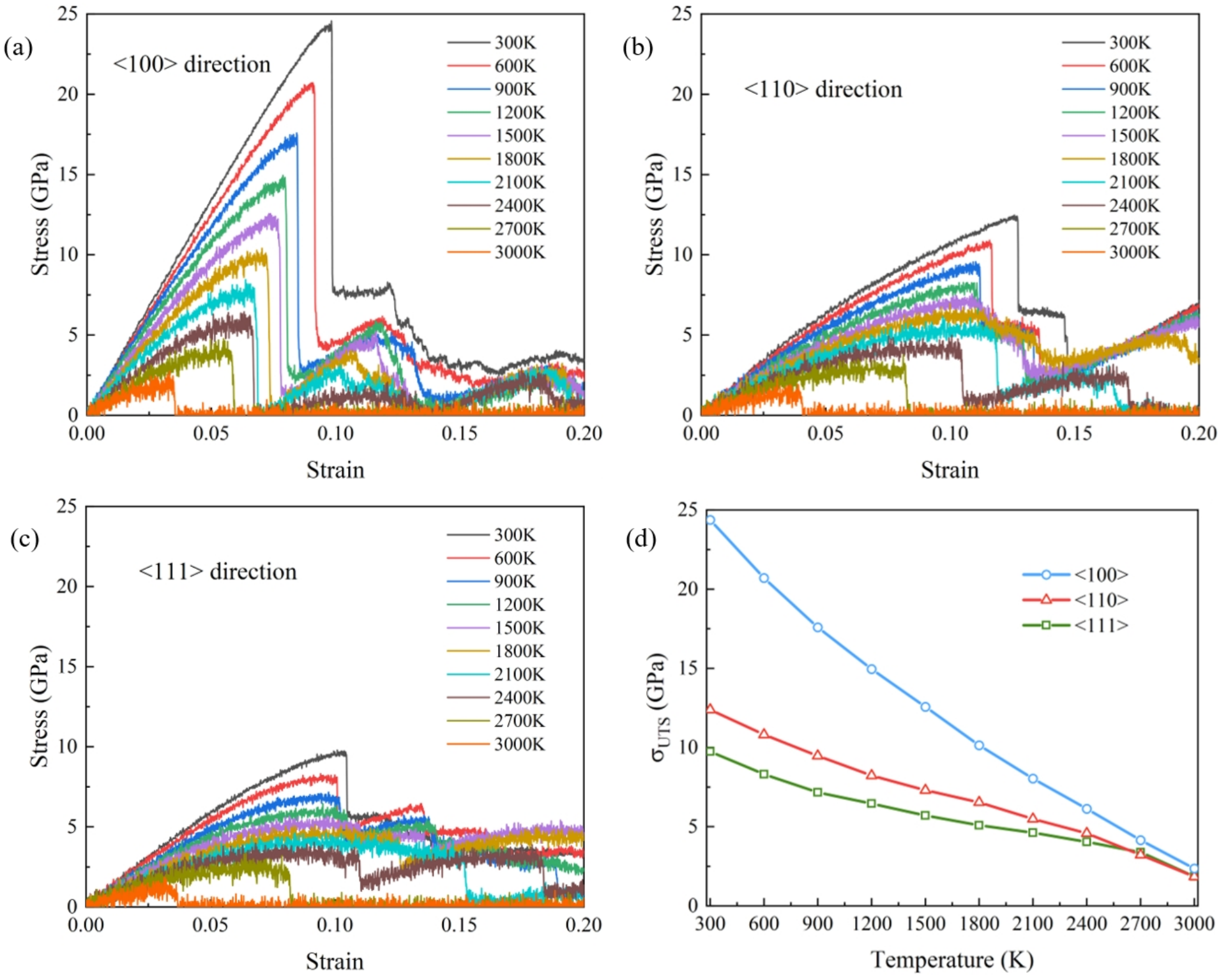

**Fig. 4** Uniaxial tensile stress-strain curves of single-crystalline $UO_2$ along the (a) <100>, (b) <110>, and (c) <111> directions, along with (d) the corresponding ultimate tensile strength ($\sigma_{UTS}$) variations as a function of temperature from 300 to 3000K.

Following the analysis of the elastic softening, uniaxial tensile simulations are conducted to assess the temperature-dependent mechanical response of $UO_2$. Figs. 4(a)-(c) present the stress-strain curves of single-crystalline $UO_2$ along the <100>, <110>, and <111> loading directions with increasing temperature. These tensile simulations are conducted at a typical strain rate of $10^8$ $s^{-1}$ under the isothermal-isobaric (NPT) ensemble with the Martyna-Tuckerman-Tobias-Klein (MTTK) integrators [74]. As shown in Figs. 4(a)–(c), the uniaxial tensile responses of single-crystalline $UO_2$ are

`

highly sensitive to both the loading directions and temperatures. Specifically, at lower temperatures, the stress-strain curves exhibit the typical brittle fracture behavior characteristic. The ultimate tensile strength ($\sigma_{UTS}$), denoting the material's upper load-bearing limit before fracture, can be determined by extracting the peak stresses from the stress-strain curves. Evidently, the $\sigma_{UTS}$ along the <100> direction (24.4 GPa) is significantly higher than those along the <110> (12.3 GPa) and <111> directions (9.7 GPa) at 300K, which can be attributed to the differences in chemical bond strength and atomic packing density along these crystallographic directions in such fluorite-type material. As clearly provided in Fig.4(d), the $\sigma_{UTS}$ exhibits a continuous and near-linear decrease with the temperature increasing across all three crystallographic orientations. Despite the <100> direction remains the strongest orientation of single-crystalline $UO_2$, the orientation dependence of $\sigma_{UTS}$ gradually diminishes as the temperature increases. Unlike the pronounced differences observed at 300 K, the $\sigma_{UTS}$ along these three crystallographic directions converges to a unified value of approximate 2 GPa when the temperature exceeds 2700 K. This vanishing anisotropy is primarily driven by the substantial lattice expansion and anharmonicity at elevated temperatures. As the temperature approaches the melting point, intense thermal fluctuations overwhelm the directional bonding and deteriorate the overall structural integrity, ultimately pushing the material toward the lower bound of its mechanical stability. Notably, the orientational anisotropy and temperature-dependent uniaxial tensile behaviors, as captured by our NEP model, align remarkably well with prior theoretical modeling [17, 75, 76] and experimental observations [77, 78].

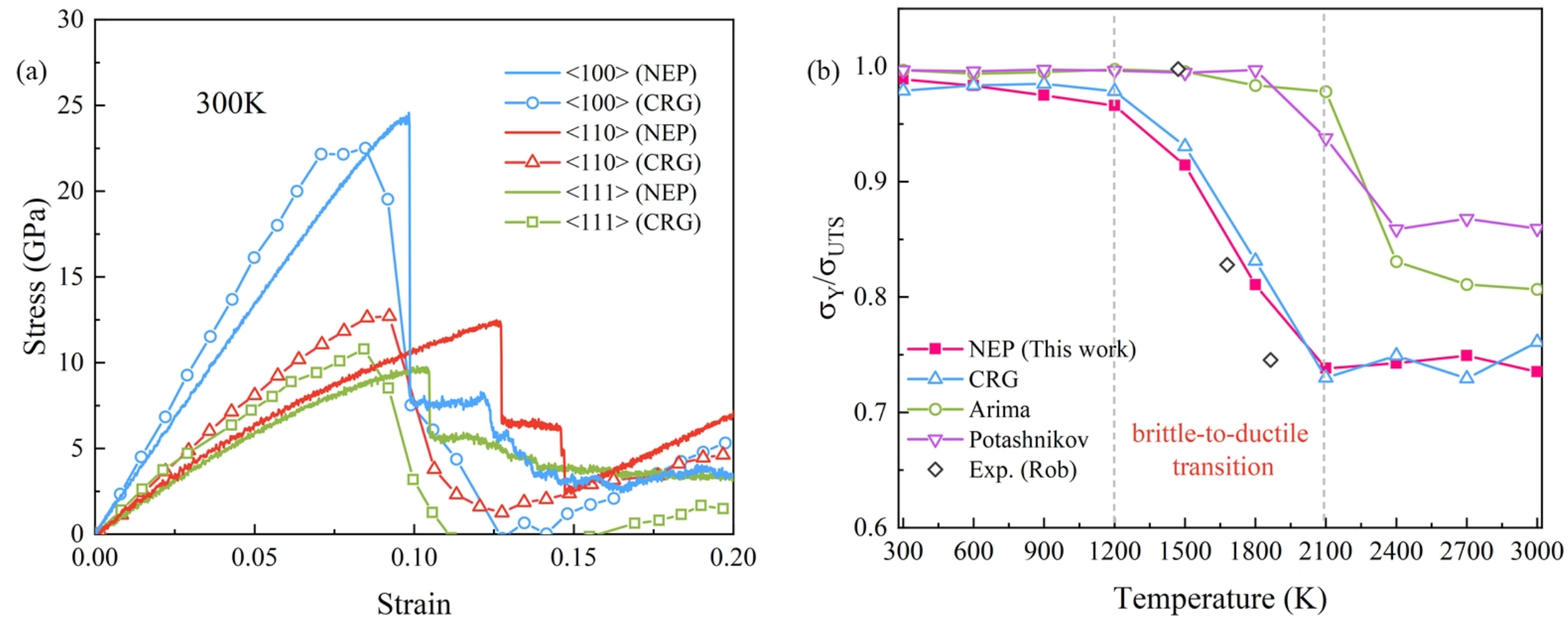

**Fig. 5** (a) The comparisons of the uniaxial tensile stress-strain curves for single-crystalline $UO_2$ along the <100> (blue), (b) <110> (red), and (c) <111> (green) directions between the NEP model and reference CRG potential at 300K. (b) Temperature dependence of the yield-to-ultimate strength ratio ($\sigma_Y / \sigma_{UTS}$) along the <111> direction. The region demarcated by the vertical dashed lines highlights the onset of brittle-to-ductile transition (BDT) predicted by our NEP model compared aganist other EIPs and experimental measurements.

Fig.5(a) further compares the uniaxial tensile stress-strain curves of single-crystalline $UO_2$ along the <100>, <110>, and <111> loading directions at 300K, as predicted by the NEP and the reference CRG potentials. Clearly, the $\sigma_{UTS}$ predicted by the NEP is slightly higher than that of CRG potential along the <100> direction, while they are remarkably close to each other along the <110> and <111> directions. However, a delay in the occurrence of fracture across all three directions is observed for the NEP compared with the CRG potential. This delayed fracture indicates that the NEP predicts a higher capacity for strain energy storage prior to crack nucleation. Despite these minor deviations in $\sigma_{UTS}$ and fracture strain, the brittle fracture behavior and pronounced mechanical anisotropy are successfully captured by our NEP, consistent with the results of the CRG potential. Beyond the low-temperature fracture behavior, precisely predicting the brittle-to-ductile transition (BDT) as temperature increases is of paramount importance for assessing the thermo-mechanical response and structural integrity of $UO_2$ at elevated temperatures. As inferred from Fig. 5, the <111> direction is the most susceptible to fracture under equal stresses. Therefore, we focus our BDT assessment exclusively on uniaxial loading along the <111> direction. To accurately determine the yield strength ($\sigma_Y$), cyclic load-unload simulations are conducted on the single-crystalline $UO_2$. Specifically, uniaxial tensile loading is performed along the <111> direction at a strain rate of $10^8$ $s^{-1}$. The yield strength ($\sigma_Y$) is then extracted at the onset of irreversible plastic deformation, corresponding to the exact point where a non-zero residual strain persists upon fully unloading the system to a zero-stress state. In the low-temperature regime ($T \leq 1200$ K), the single-crystalline $UO_2$ exhibits purely brittle behavior characterized by zero residual strain prior to fracture. Consequently, the yield strength ($\sigma_Y$) is nearly identical to the ultimate tensile strength ($\sigma_{UTS}$), resulting in a $\sigma_Y / \sigma_{UTS}$ ratio of approximately unity. As the temperature enters the intermediate regime ($1200 \text{ K} \leq T \leq 2100 \text{ K}$), a sharp decline in the $\sigma_Y / \sigma_{UTS}$ ratio is observed, signifying the onset of the BDT. At temperatures above 2100K, the $\sigma_Y / \sigma_{UTS}$ ratio stabilizes at a plateau value of approximately 0.75, corresponding to a fully ductile state characterized by stable plastic flow prior to fracture. Notably, the BDT temperature and high-temperature $\sigma_Y / \sigma_{UTS}$ ratio predicted by our NEP are lower than those derived from the Arima [79] and Potashnikov potentials [80], while exhibiting significantly closer agreement with the CRG potential and experimental measurements [81]. Consdering that the CRG potential is a well-established benchmark for $UO_2$ mechanics due to its faithful reproduction of the Cauchy violation [14, 17], the excellent agreement between our NEP's and the CRG potential's results firmly validates the reliability of the NEP in describing the mechanical behavior of $UO_2$.

**3.2 Energetics of point and extended defects**

In addition to mechanical properties, accurately describing the defect behaviors is essential for modeling defect-mediated species diffusion and irradiation damage of $UO_2$. Firstly, the formation energies of primary point defects are calculated utilizing our NEP.

`

To minimize periodic image interactions, a 3×3×3 single-crystalline $UO_2$ supercell containing 324 atoms is employed as the reference crystal. Full geometry optimization is performed on both the reference and defect-containing configurations using the FIRE optimization algorithm implemented within the ASE package. To accommodate defect-induced local volume changes under simulated zero-pressure conditions, the ExpCellFilter class is utilized to simultaneously relax the atomic coordinates and cell vectors. Finally, the structural minimization is considered converged when the maximum residual force falls below 0.01 eV/Å. To satisfy the charge neutrality requirement within $UO_2$, only the formation energies of charge-balanced point defect pairs are considered. In total, two main types of point defects are systematically investigated: (1) Schottky defects (SDs), comprising neutral tri-vacancy clusters (1 $V_U^{''''}$ +2 $V_O^{\cdot\cdot}$ ). These can be divided into isolated SDs ($SD_{isolated}$) containing non-interacting vacancies, and bound SDs (BSDs) comprising tightly clustered vacancies. BSDs are further categorized into edge-sharing ($BSD_1$), face-sharing ($BSD_2$), and diagonal ($BSD_3$) configurations, corresponding to the oxygen vacancies located at the first-($1^{st}$-NN), second-($2^{nd}$-NN), and third-nearest neighbor ($3^{rd}$-NN) sites on the anion sublattice, respectively. (2) Frenkel pairs (FPs), defined as associated vacancy-interstitial pairs. Based on their separation distances, these are classified into non-interacting isolated FPs (O-$FP_{isolated}$, U-$FP_{isolated}$), and stable closely bound FPs. Due to spontaneous recombination of $1^{st}$- and $2^{nd}$-NN pairs during relaxation, only $3^{rd}$-NN (O-$FP_{3rd}$, U-$FP_{3rd}$) and $4^{th}$-NN (O-$FP_{4th}$, U-$FP_{4th}$) configurations are employed for calculating their formation energy barriers.

The formation energies ( $E_f$ ) are calculated as the total energy difference between the defect-containing and fully relaxed reference supercells. For the SD defects, the formation energy is derived from:

$$E_f(SD) = E_{SD} - \frac{N-1}{N} E_{reference} \tag{12}$$

where $N$ is the number of uranium ions in the perfect supercell ($N$=108 in this study), $E_{SD}$ is the total energy of the supercell containing Schottky defects, and $E_{reference}$ is the total energy of reference supercells.

Regarding the FPs defects, $E_f(FP)$ is calculated as:

$$E_f(FP) = E_{FP} - E_{reference} \tag{13}$$

where $E_{FP}$ and $E_{reference}$ are the total energies of the FPs-containing and perfect reference supercells, respectively.

**Table.2** Comparisons of the formation energies for various point defects in $UO_2$ calculated by the NEP against those by other interatomic potentials, DFT+U calculations and experimental measurements. The investigated point defects include the isolated ($SD_{isolated}$) and bound Schottky defects (edge-sharing $BSD_1$, face-sharing $BSD_2$, and diagonal $BSD_3$), alongside isolated ($FP_{isolated}$) and specific neighbor-separated ($3^{rd}$- and $4^{th}$-NN) oxygen and uranium Frenkel pairs ($FP_{3rd}$, $FP_{4th}$). Note that the dash ("-") indicates no available value.

| Defect type | NEP (This work) | MLIP-DFT+U [10] | SNAP [22, 66] | HDNNP [22] | CRG | SMTB-Q | DFT+U [8, 82, 83] | Exp. [84] |
|---|---|---|---|---|---|---|---|---|
| $SD_{isolated}$ | 5.38 | 5.28 | - | - | 10.64 | - | 4.2-11.8 | 6-7 |
| $BSD_1$ | 3.66 | 3.52 | 4.09 | 4.03 | 6.291 | 4.27 | 3.32-4.24 | - |
| $BSD_2$ | 3.36 | 3.07 | 3.08 | 3.55 | 5.22 | 3.51 | 2.54-3.93 | - |
| $BSD_3$ | 3.47 | 3.17 | 3.23 | 3.75 | 5.14 | 3.59 | 2.82-3.89 | - |
| O-$FP_{isolated}$ | 4.12 | 4.69 | 4.66 | 4.08 | 5.73 | - | 2.4-7.0 | 3-4 |
| O-$FP_{3rd}$ | 3.95 | 3.86-3.89 | 4.06 | - | 0.0 | 2.80 | 2.4 | - |
| O-$FP_{4th}$ | 4.01 | 3.86-3.89 | 4.3 | - | 5.66 | 4.53 | - | - |
| U-$FP_{isolated}$ | 10.94 | 9.47 | - | - | 15.47 | - | 9.1-16.5 | 9.5 |
| U-$FP_{3rd}$ | 10.39 | 6.53 | 9.21 | - | 10.91 | 5.54 | 9.1 | - |
| U- $FP_{4th}$ | 10.47 | - | 11.72 | - | 12.05 | 5.66 | - | - |

`

As clearly seen in Table 2, the formation energies of almost all defects calculated by the NEP agree with those derived from other MLIPs (MLIP-DFT+U, SNAP, and HDNNP) and DFT+U calculations. In comparison to the CRG potential, which is reported to slightly overestimate these values [85], our NEP yields lower formation energies that are more consistent with experimental data. More importantly, the NEP accurately captures not only the absolute values of defect formation energies but also their relative differences. Specifically, the formation energy of isolated Schottky defects ($SD_{isolated}$) is notably higher than those of bound Schottky defects (BSDs), reflecting a strong tendency for SDs to stabilize into bound clusters. Analogous to the SDs, both uranium and oxygen FPs exhibit a clustering tendency, with bound configurations ($FP_{3rd}$ and $FP_{4th}$) being energetically favorable over their isolated counterparts. Additionally, the formation energies of U-related defects are significantly higher than those of O-related defects, aligning with the reported findings [85].

To systematically evaluate the defect migration kinetics, the distinct diffusion pathways for oxygen and uranium vacancies ($V_O^{\bullet\bullet}$, $V_U^{''''}$) and interstitials ($O_i^{''}$, $U_i^{\bullet\bullet\bullet\bullet}$) are studied. For the vacancies, migration pathways are modeled by considering direct hopping along the primary crystallographic directions (i.e., <100>, <110>, and <111>). The lowest energy barrier among these pathways is identified as the migration energy for the respective vacancy. Specifically, the initial state (IS) is created by introducing a single vacancy into the perfect supercell, while the final state (FS) is generated by displacing a specific neighboring O or U ion from the target coordination shell into this vacant site. For the interstitials, the IS is constructed by placing an interstitial atom at an octahedral site within the perfect supercell. Regarding oxygen interstitials ($O_i^{''}$), the transition to the FS is governed by two distinct mechanisms. The first pathway involves a direct hop to an adjacent octahedral site. The second is a collinear <100> interstitialcy (knock-on) mechanism, where the migrating $O_i^{''}$ displaces a lattice O-ion into its neighboring interstitial position. Uranium interstitials ($U_i^{\bullet\bullet\bullet\bullet}$) are also evaluated along their optimal migration paths. With the IS and FS established for each defect migration pathway, the minimum energy paths (MEPs) and their corresponding transition states (TS) are determined using the Climbing Image Nudged Elastic Band (CI-NEB) method implemented in ASE package. To provide a robust initial guess for the migration trajectory, the pathways are first generated via Image Dependent Pair Potential (IDPP) interpolation, utilizing 27 intermediate images. The CI-NEB optimization is subsequently performed to drive the highest-energy image exactly to the saddle point, ensuring precise TS identification. The structural relaxation along the band is considered converged when the maximum residual force falls below 0.05 eV/Å. Finally, the migration energy barrier ( $E_m$ ) is calculated as the energy difference between the converged saddle point and fully relaxed IS:

$$E_m = E_{TS} - E_{IS} \tag{14}$$

where $E_{TS}$ and $E_{IS}$ denote the total potential energies of TS and IS of the system, respectively. The resulting energy profiles along the reaction coordinate, derived from these CI-NEB calculations, are plotted in Fig. 6. These profiles provide a direct visualization of the MEPs for all investigated point defects along distinct pathways.

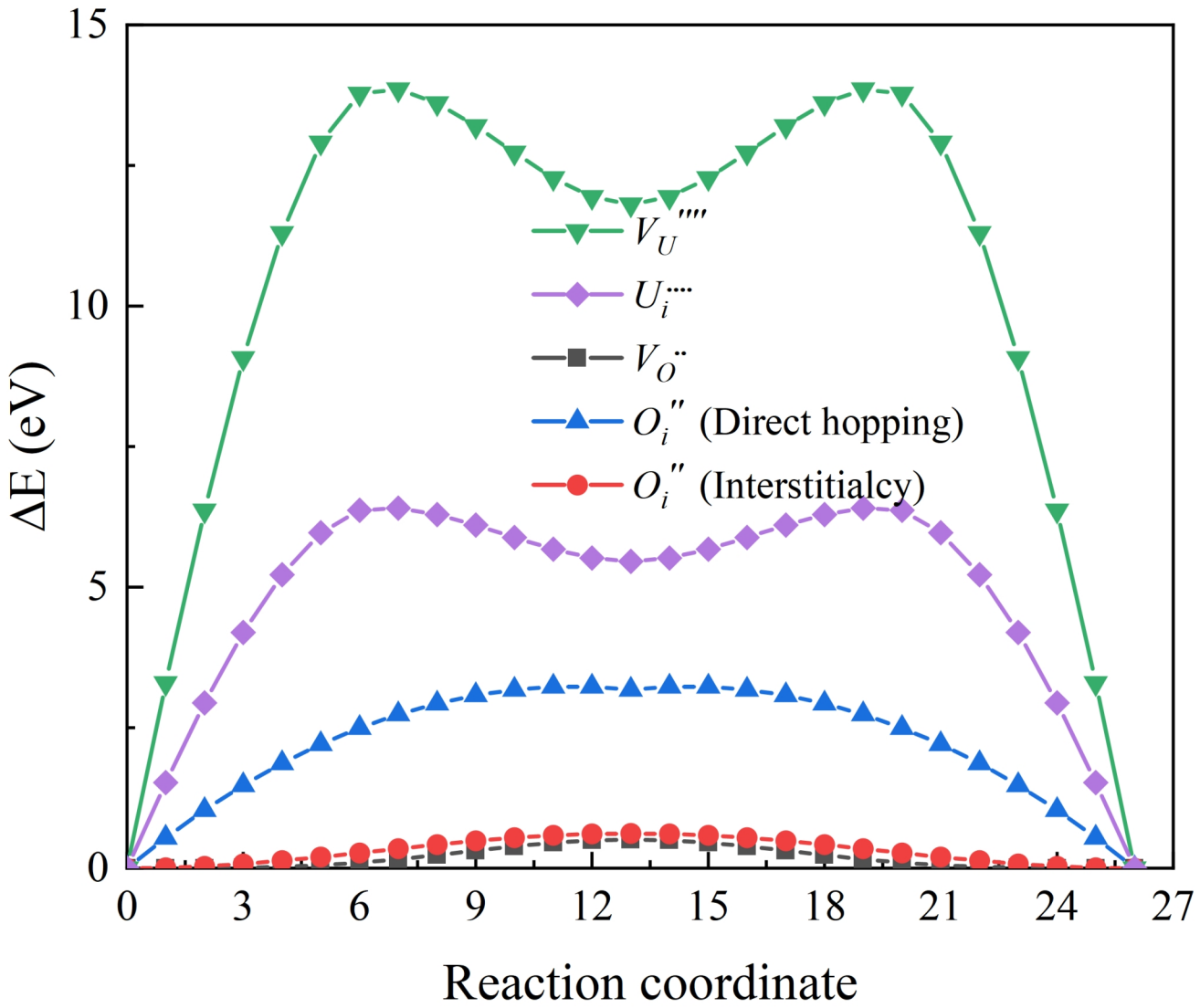

**Fig. 6**. Migration energy profiles of various point defects in $UO_2$ calculated by the NEP. The energy variations along the reaction coordinate are illustrated for uranium

`

vacancies ($V_U^{''''}$), uranium interstitials ($U_i^{\bullet\bullet\bullet\bullet}$), oxygen vacancies ($V_O^{\bullet\bullet}$),and oxygen interstitials ($O_i^{''}$) migrating via direct hopping and interstitialcy mechanisms.

**Table.3** Comparisons of the migration energies for various point defects in $UO_2$ calculated by the NEP against those by other interatomic potentials, DFT/DFT+U calculations, defect simulations, and experimental measurements. The investigated migration pathways include oxygen and uranium vacancies ($V_O^{\bullet\bullet}$ and $V_U^{''''}$), as well as oxygen and uranium interstitials ($O_i^{''}$ and $U_i^{\bullet\bullet\bullet\bullet}$). Note that the dash ("-") indicates no available value.

| Defect type | NEP (This work) | CRG [86] | Arima [87] | Yakub [13] | Bai (TAD) [88] | DFT+U (Dorado) [89] | DFT+U (Gupta) [90] | DFT (Yun) [91] | Defect simulations (Jackson) [92] | Exp. [84] |
|---|---|---|---|---|---|---|---|---|---|---|
| $V_O^{\bullet\bullet}$ | 0.51 | 0.24-3.82 | 1.5 | 1.3 | 0.29 | 0.67 | 1.01 | 0.63 | 0.53 | 0.7-1.5 |
| $O_i^{''}$ (Interstitialcy) | 0.62 | 0.65 | - | - | 1.0 | 0.81-0.93 | 1.13 | - | 0.64 | 0.8-1.0 |
| $O_i^{''}$ (Direct hopping) | 3.23 | - | - | - | - | 3.22 | 2.14 | - | - | - |
| $V_U^{''''}$ | 13.85 | 4.34-16.48 | 5.6 | 1.9 | - | - | - | 3.09 | 6.0 | 2.4 |
| $U_i^{\bullet\bullet\bullet\bullet}$ | 6.40 | 7.13 | - | - | - | - | - | - | 8.76 | - |

As shown in Fig. 6 and Table 3, the NEP-predicted migration energies for various point defects in $UO_2$ are systematically compared with reported literature data, encompassing

`

other EIPs, DFT, defect simulation, and experimental measurements. For the highly mobile O-ions, the NEP model predicts a migration barriers of 0.51 eV for $V_O^{\bullet\bullet}$ and 0.62 eV for $O_i^{''}$ via the interstitialcy mechanism. These values exhibit excellent agreement with theoretical defect simulations (0.53 eV and 0.64 eV, respectively) and align closely with the ranges of DFT calculations. Besides, the NEP model predicts a migration energy of 3.23 eV for $O_i^{''}$ via the direct hopping, which is nearly identical to Dorado's DFT+U data (3.22 eV). Our NEP accurately reproduces the significant energy difference between the interstitialcy and direct hopping mechanisms of $O_i^{''}$ migration, confirming that the interstitialcy process is energetically more favorable, as reported in the literatures [88, 89, 93]. Also, the migration energy of $V_O^{\bullet\bullet}$ is lower than that of $O_i^{''}$, which is consistent with experiments [94] and DFT studies [89]. With respect to the far less mobile U-ions, their migration barriers are substantially higher (13.85 eV for $V_U^{''''}$ and 6.4 eV for $U_i^{\bullet\bullet\bullet\bullet}$). Furthermore, the pronounced gap between the migration energies of uranium and oxygen defects, fundamentally reflects the distinct structural stabilities of these two sublattices. Collectively, these comparative evaluations of both formation and migration energies demonstrate the ability of the NEP to accurately reproduce the profound disparity in the structural evolution and migration kinetic between the uranium and oxygen sublattices.

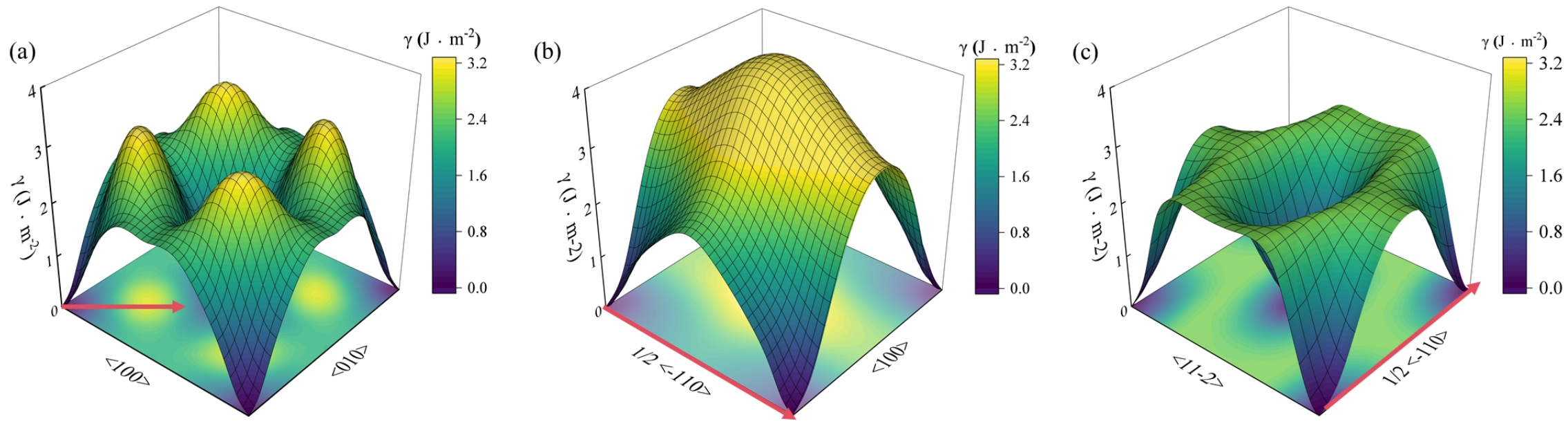

**Fig. 7** Generalized stacking fault energy surfaces ( $\gamma$ surfaces) of the (a) {100}, (b) {110}, and (c) {111} slip planes in $UO_2$, as predicted by the NEP. The Burgers vector, corresponding to the 1/2<110> slip direction, is marked by a red arrow in each plane.

Generalized stacking fault (GSF) energy surfaces, commonly referred to as $\gamma$ surfaces, serve as an indispensable prerequisite for accurately modeling dislocation mobility and plastic deformation. Therefore, to rigorously validate the predictive capability of the NEP for extended defects, we systematically examine the $\gamma$ surfaces across various representative slip systems in $UO_2$. The γ surfaces are examined for the {100}, {110}, and {111} slip planes, which host the primary dislocations observed in experiments [95]. To construct the $\gamma$ surfaces for representative {100}, {110}, and {111} slip planes, suitably oriented supercells are constructed with the slip plane aligned perpendicular to the x-axis. The GSFE is mapped by rigidly translating the upper half of the supercell relative to the lower half across a $40\times40$ grid of displacement vectors $\mathbf{u}=(u_y,u_z)$ in the yz-plane. Atomic positions are only allowed to relax along the x-direction, preventing the system from spontaneously reverting to its original state. Under PBCs, this uniform translation inevitably introduces two equivalent stacking fault interfaces within the simulation box. Therefore, the γ-surface energy is calculated by:

$$\gamma(\mathbf{u})=\frac{E(\mathbf{u})-E(0)}{2A} \tag{15}$$

where $E(\mathbf{u})$ and $E(0)$ are the total energies of the faulted supercell with displacement vector $\mathbf{u}=(u_y,u_z)$ and the fully relaxed perfect supercell, respectively, and $A$ represents the cross-sectional area of the slip plane. Fig. 7 presents the three-dimensional $\gamma$ surfaces for the {100}, {110}, and {111} slip planes mapped using the NEP. Fundamentally, these topographical energy landscapes determine the mobility of extended defects, thereby governing the plastic deformation behaviors of $UO_2$. Notably, the three-dimensional $\gamma$ surfaces predicted by our NEP agree well with those reported by Dubois *et al*, which include DFT+U calculations, alongside their SNAP and HDNNP potentials [22]. Besides, the lowest-energy valleys traversing these surfaces dictate the minimum energy paths (MEPs) for dislocation glide, with the most energetically favorable slip trajectories (Burgers vectors) explicitly marked by red arrows (Fig. 7).

`

More discussions of the three-dimensional $\gamma$ surfaces are provided in Section S7 of the Supplementary Materials.

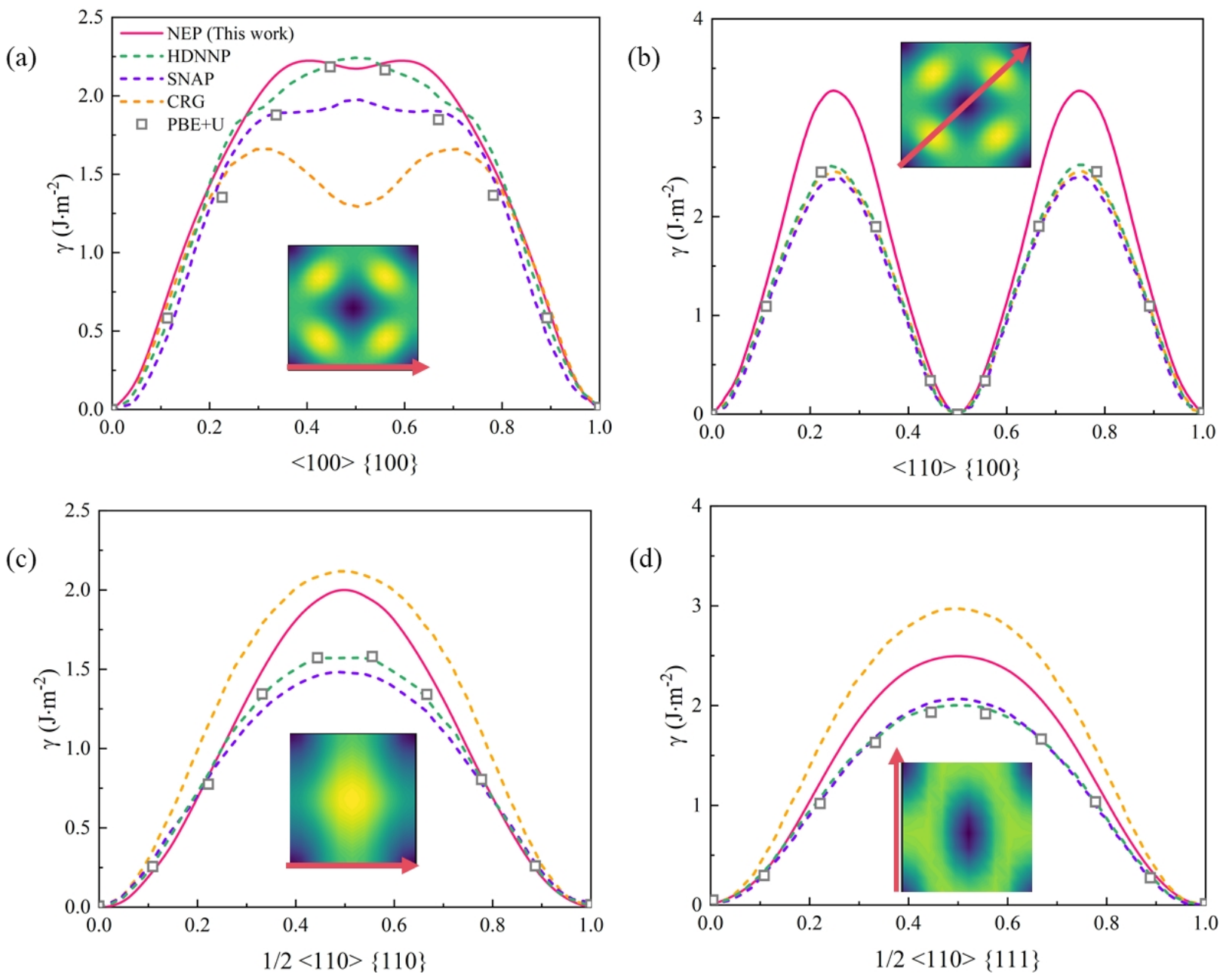

**Fig. 8** One-dimensional generalized stacking fault energy ($\gamma$) profiles for the representative slip systems in $UO_2$: (a) <100>{100}, (b) <110>{100}, (c) 1/2<110>{110}, and (d) 1/2<110>{111}. The predictions by the NEP are compared against those by other interatomic potentials and PBE+U calculations. Insets illustrate the corresponding two-dimensional $\gamma$-surfaces, where the red arrows mark the specific slip directions along which the one-dimensional profiles are extracted.

To further quantitatively validate the predictive accuracy of the NEP regarding the Generalized stacking fault (GSF) energy of $UO_2$, one-dimensional energy profiles are extracted along these MEPs and compared with reference data shown in Fig. 8. For the {100} slip plane, atomic glide preferentially occurs along the <100> slip direction (Fig. 8a), which exhibits a significantly lower energy barrier (~2.2 J/m$^2$) compared to the <110> slip direction (~3.2 J/m$^2$) (Fig. 8(b)). Clearly, for these slip directions within the {100} plane, the overall profiles of the MEPs predicted by our NEP show close agreement with those from other potentials and PBE+U calculations. Specifically, the energy barrier $\gamma$ in the <100> direction exhibits excellent agreement with that calculated from the HDNNP potential and PBE+U (Fig. 8(a)). Regarding the <110> slip direction, although the NEP-predicted $\gamma$ is higher than the DFT calculations, they remain within a physically reasonable range (Fig. 8(b)). Moreover, the NEP-predicted $\gamma$ values are lower than the CRG predictions on the {110} (Fig. 8(c)) and {111} (Fig. 8(d)) planes, which corresponds to the well-documented tendency of the CRG potential to overestimate these values [22]. In summary, these comparative analyses confirm the high fidelity of the NEP in accurately capturing the energetics of both point and extended defects.

### 3.3 Thermophysical quantities

To accurately determine the melting temperature ( $T_m$ ) of $UO_2$ using the NEP, two-phase simulation (solid-liquid coexistence) method is employed. Initially, an 8×8×16 $UO_2$ supercell containing 12,288 atoms is constructed. To establish the two-phase coexistence system, the initial uniform configuration is first equilibrated at 2200 K under the NPT ensemble (0 atm.) for 1 ns to fully release residual internal stresses, followed by an additional 500-ps relaxation under the NVT ensemble. At this stage, the entire system is in a solid state (Fig. 9(a)), with the uranium sublattice fully identified as a face-centered cubic (FCC) structure via the CNA method (Fig. 9(c)). Subsequently, the simulation box is spatially partitioned into two equal halves. The atoms in one half are positionally constrained to preserve the solid crystalline structure, while those in the unconstrained half are heated to 5000 K under NPT ensemble (0 atm.) for 1ns to induce complete melting. To minimize the interfacial stress mismatch between the solid and liquid phases, the unconstrained liquid region is subsequently relaxed at 3600 K (slightly above the expected $T_m$ of $UO_2$) for 1 ns [96]. Following this relaxation, the positional constraints on the solid region are released, bringing the solid phase (2200 K) and liquid phase (3600 K) into direct contact. This procedure successfully establishes a stable solid-liquid coexistence configuration (Fig. 9(b)), where the uranium sublattice is identified as comprising the intact FCC and disordered ('Other')

`

phases, respectively (Fig. 9(d)). Using this established configuration, a series of production NPT simulations are performed over a total duration of 37 ns, spanning a temperature range from 300 K to 3600 K at 100K intervals and zero pressure. All thermodynamic quantities are derived by averaging over the final 200 ps of the MD trajectories at each temperature step. The occurrence of melting is signified visually by the migration of the solid-liquid interface, and quantitatively by sharp discontinuities in the temperature-dependent density and enthalpy increment ( $H(T)-H(300K)$ ). As shown in Fig. 9(e) and 9(f), the onset of melting is quantitatively evidenced by abrupt discontinuities in both density and enthalpy increment ( $H(T)-H(300K)$ ) curves at 3125 K. Furthermore, the corresponding insets visually corroborate this melting process, capturing the complete structural transition from a highly ordered solid into a disordered liquid phase. This transition is characterized by both the global amorphization of the overall lattice (Fig. 9(f) insets) and the structural collapse of the FCC uranium sublattice (Fig. 9(e) insets). Notably, compared to the DP and CRG potentials, the density and enthalpy increment values predicted by our NEP are nearly identical to the experimental data recommended by Fink [65]. More crucial, the predicted $T_m$ of 3125 K is in excellent agreement with the experimental estimation of 3120 K [23, 97], yielding a negligible deviation of only 0.16%. Successfully reproducing the $T_m$ of $UO_2$ with such high precision provides our strong confidence for further investigations into its broader thermodynamic properties, particularly at elevated temperatures.

`

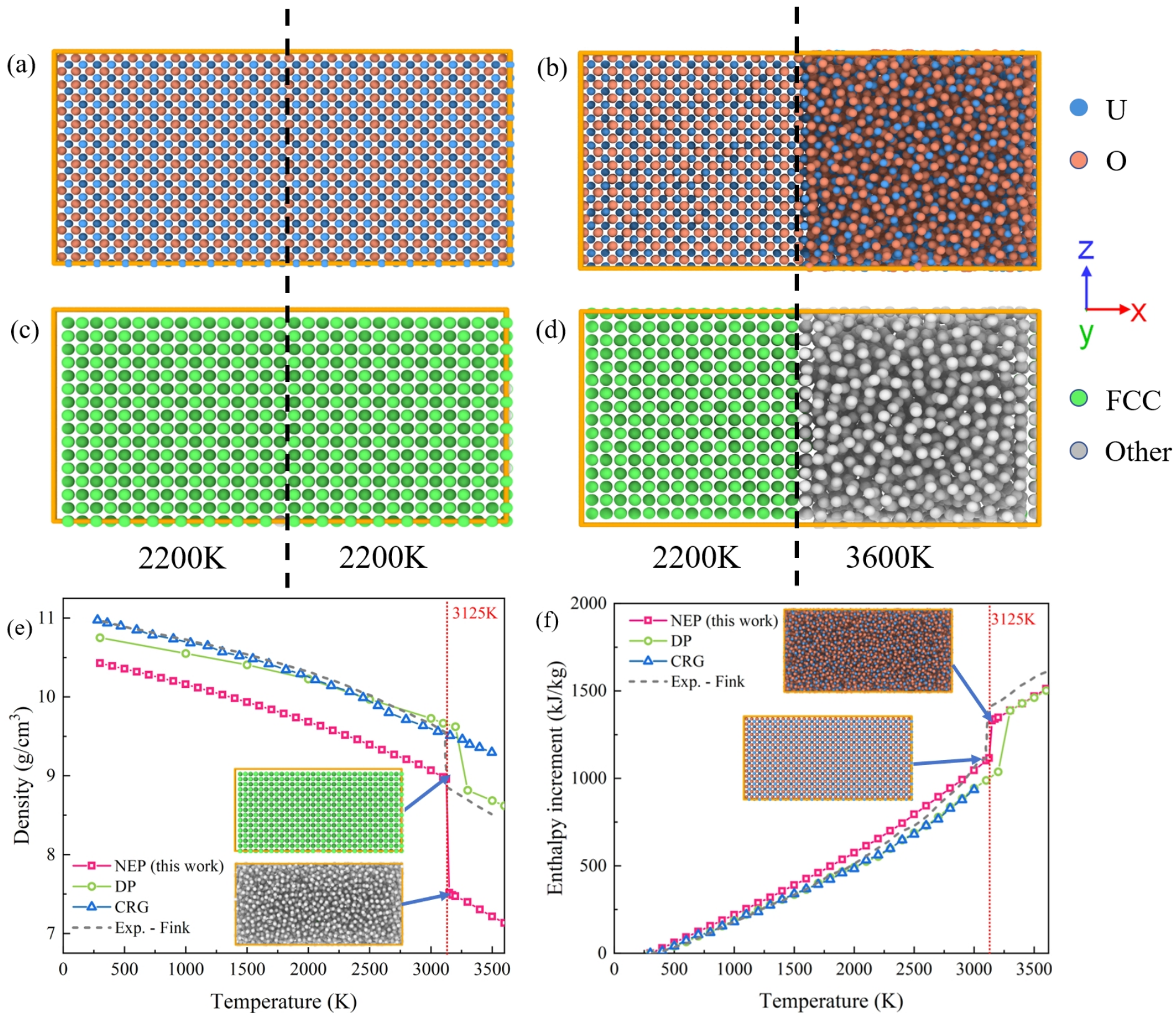

**Fig. 9**. Two-phase simulation (TPS) modeling of the $UO_2$ melting behavior. Atomic configurations (a, b) and their corresponding Common Neighbor Analysis (CNA) identification of the U sublattice (c, d) for solid-solid and solid-liquid coexistence models. For better visualization, CNA is only performed on the uranium sublattice. Blue/red spheres denote U/O ions, while green/white colors represent the crystalline FCC and disordered liquid phases identified via CNA, respectively. Temperature-dependent variations of the (e) density and (f) enthalpy increment ($H(T)-H(300K)$). The predictions of the NEP are compared with those of other potentials (DP, CRG) and Fink's recommended experimental data. A sharp discontinuity is observed at 3125 K (red dotted line), signifying the onset of melting, while the insets capture the corresponding microstructural change across this estimated melting temperature.

To investigate the temperature-dependent thermal expansion, long-duration heating simulations are conducted under the NPT ensemble with the SCR barostat from 300 K to 3600 K at 100 K intervals. For each temperature step, a continuous 1 ns simulation (comprising 500 ps for structural relaxation and 500 ps for production sampling) is conducted on the single-crystalline $UO_2$ supercell. The temperature-dependent equilibrium lattice constant ( $a_T$ ) is subsequently determined by time-averaging the simulation box dimensions over the final 200 ps of the production stage for each sampled temperature.

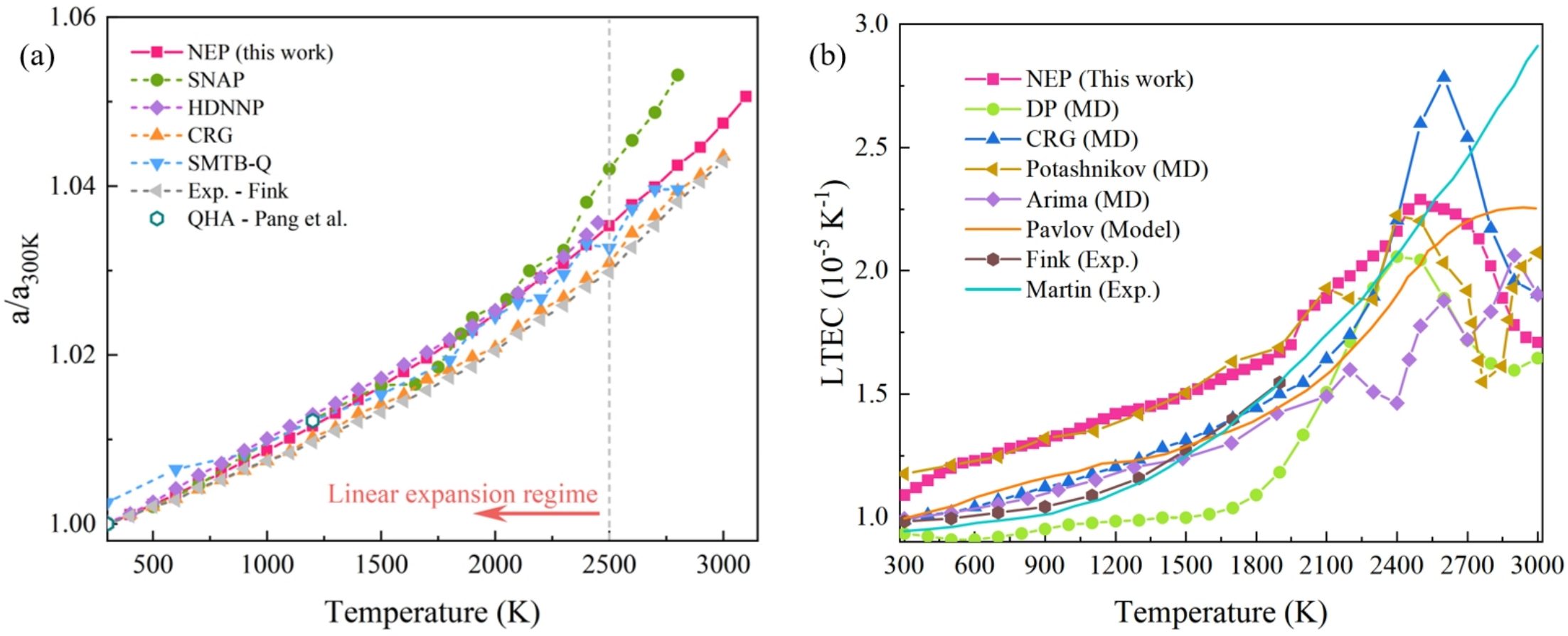

**Fig. 10**. Comparisons of the temperature-dependent thermal expansion properties of $UO_2$ predicted by the NEP against available theoretical calculation and experimental data. (a) Variations of the relative equilibrium lattice constant ( $a_T / a_{300K}$ ) from 300 K to 3100 K. The linear expansion regime is marked by a red arrow, with a vertical dashed line indicating the onset of non-linear structural expansion at elevated temperatures. (b) The corresponding linear thermal expansion coefficient (LTEC) variations as a function of temperature.

Fig. 10(a) compares the temperature-dependent variations of $a_T / a_{300K}$ calculated by the NEP against other interatomic potentials, first principle QHA calculations [98], and experimental measurements [65]. Clearly, the $a_T / a_{300K}$ values calculated by our NEP fall within an intermediate range bounded by the highest (SNAP potential) and the lowest (CRG potential and Fink's experimental data) values, while aligning closely with the HDNNP, SMTB-Q, and first-principles QHA calculations. Moreover, the temperature-dependent variations of the NEP-calculated $a_T / a_{300K}$ is highly consistent with other reported data, successfully capturing a pronounced transition near 2500 K [17]. Below 2500 K, the $a_T / a_{300K}$ increases almost linearly, exhibiting a linear thermal expansion regime. However, a significant non-linear expansion emerges once the temperatures exceed 2500 K. This noticeable deviation from linearity stems from an anomalous volume expansion, which is induced by the extensive accumulation of oxygen Frenkel defects during the superionic transition [19].

To quantitively characterize this transition in thermal expansion behavior, the temperature dependence of the linear thermal expansion coefficient (LTEC, $\alpha_L$ ) is further examined. The $\alpha_L$ is derived by calculating the isobaric first-order derivative of the lattice parameter versus temperature as:

$$\alpha_L = \frac{1}{a_{300K}} \left( \frac{\partial a_T}{\partial T} \right)_P \tag{16}$$

where $a_{300K}$ denotes the reference lattice parameter at 300 K, and the temperature derivative $\partial a_T / \partial T$ is obtained via numerical differentiation. The calculated LETC exhibits a steady monotonic rise in the low-temperature regime, subsequently generating a pronounced "λ-shape" peak at elevated temperatures (Fig.10(b)). This characteristic "λ-peak" spans several hundred Kelvins and serves as a signature of the superionic transition (Bredig) transition, which is due to the abrupt structural evolution induced by the pre-melting of oxygen sublattice [17, 47, 99]. As predicted by the NEP, the LTEC "λ-peak" occurs within a temperature range of 2400~2700 K, corresponding to 0.77~0.86 $T_m$ ($T_m$ = 3125 K). This aligns closely with the widely recognized value of approximately 0.85 $T_m$ for activating the ST [28-30]. Despite being slightly higher than the theoretical [17, 23, 100] and experimental [65, 101] data at lower temperatures, our NEP-predicted LTEC values agree exceptionally with that of the Potashnikov potential. Crucially, the rounded "λ-peak" captured by the NEP perfectly reproduces the "diffuse" nature of the ST. Consequently, this confirms the robustness of the NEP in accurately reproducing the lattice expansion of $UO_2$ across all temperatures, particularly its ability to capture the anomalous expansion behavior associated with the

`

ST. To provide a more quantitative evaluation of the temperature-dependent lattice anharmonicity, the Grüneisen parameter ($\gamma$) is calculated as:

$$\gamma = \frac{\alpha_V \cdot V_m \cdot B}{C_V} \tag{17}$$

where $\alpha_V$ represents the volumetric thermal expansion coefficient ($\alpha_V = 3\alpha_L$), $V_m$ is the molar volume, and $B$ denotes the bulk modulus. Their temperature-dependent values are obtained from Fig.10(b), Fig.10(a), and Fig.3(d), respectively. Besides, $C_V$ is the isochoric heat capacity, determined by the temperature derivative of the internal energy ($U$) at constant volume as

$$C_V = (\frac{\partial U}{\partial T})_{N,V} \tag{18}$$

The NEP-predicted Grüneisen parameter ($\gamma$) exhibits a relatively stable plateau, fluctuating between approximately 2.05 and 2.15 at temperatures up to 1200 K (Fig. 11). These predictions align closely with those from the DP MLIP and several EIPs (CRG, Arima, and Catlow), while being higher than the values derived from the Yamada and Morelon EIPs.

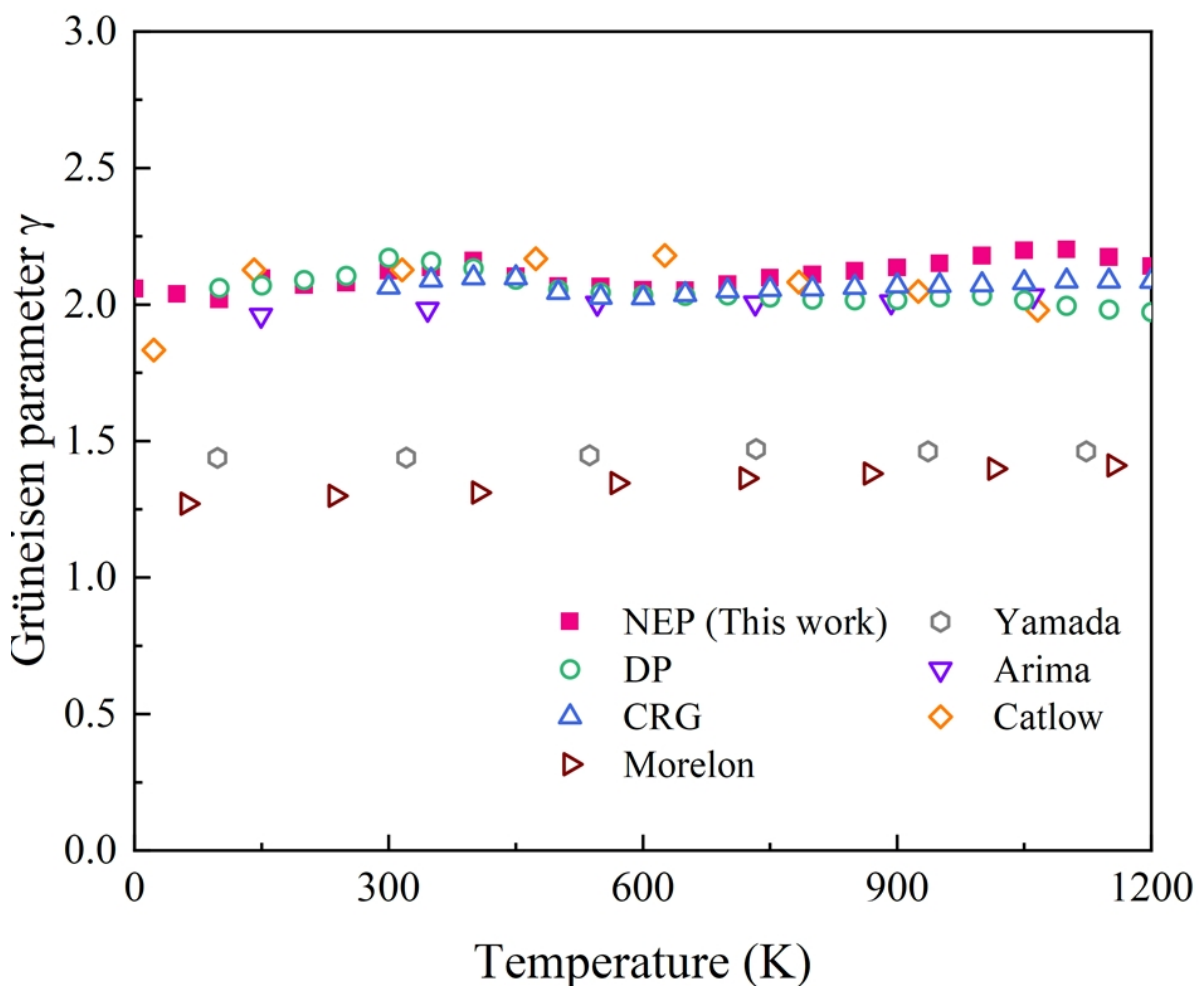

**Fig. 11**. Variations of the Grüneisen parameter ($\gamma$) as a function of temperatures. The $\gamma$ values calculated by the NEP are compared with those obtained from various EIPs (CRG, Morelon, Yamada, Arima, Catlow) and the DP MLIP.

To enable a more direct quantitative comparison with available theoretical calculations and experimental measurements, the average Grüneisen parameter ($\bar{\gamma}$) is calculated as the arithmetic mean of the temperature-dependent $\gamma$ values over the evaluated temperature range:

$$\bar{\gamma} = \frac{1}{N}\sum_{i=1}^{N}\gamma(T_i) \tag{19}$$

where $N$ represents the total number of sampled temperature points, $\gamma(T_i)$ denotes the Grüneisen parameter at a given temperature $T_i$.

**Table 4.** Comparison of the average Grüneisen parameter ($\bar{\gamma}$) calculated by the NEP against available theoretical calculations and experimental measurements. (BTE denotes the Boltzmann Transport Equation).

| Method | $\bar{\gamma}$ |
|---|---|
| NEP (This work) | 2.11 |
| DFT+U/BTE (Njifon *et al.* [102]) | 1.88 |
| Potashnikov potential/BTE (Njifon *et al.* [102]) | 1.28 |
| Exp. (Fritz *et al.* [71]) | 1.8-2.2 |
| Exp. (Yamashita *et al.* [103]) | 1.9 |
| Exp. (Serizawa *et al.* [104]) | 1.62 |
| Exp. (Willis *et al.* [105]) | 1.7 |

As summarized in Table 4, t the average Grüneisen parameter ($\bar{\gamma}$) derived by the NEP is 2.11. This value falls within the experimental range of 1.8–2.2 reported by Fritz *et al.* [71], and remains reasonably consistent with other experimental measurements (1.62–1.9) [103-105]. Furthermore, the NEP-calculated $\bar{\gamma}$ is comparable to the high-fidelity DFT+U/BTE calculation (1.88). Conversely, the $\bar{\gamma}$ predicted by the BTE using Potashnikov potential (1.28) significantly deviates from other reported values [102]. These comparisons clearly demonstrate that our NEP possesses a superior capability in capturing the intrinsic lattice anharmonicity of $UO_2$.

`

## 4. Discussions:

### 4.1 Oxygen sublattice pre-melting:

The pair distribution function (PDF) describes the probability density of finding a particle at a specific radial distance from a reference particle. Serving as a direct and fundamental measure, it accurately delineates the local structural order and coordination environment within a many-particle system. In this work, the temperature-dependent PDF is examined to evaluate the capability of the NEP in capturing the nearest-neighbor bond length variations, thermally induced lattice softening, and most importantly, the microstructural evolution during the ST. Mathematically, the like-pair PDF $g_{\alpha\alpha}(r)$, for a specific atomic species $\alpha$ (U or O) within a simulation volume $V$ comprising $N_\alpha$ atoms, can be formulated as:

$$g_{\alpha\alpha}(r) = \frac{V}{N_\alpha^2} \frac{dN_{\alpha\alpha}(r,dr)}{4\pi r^2 dr} \tag{20}$$

where $dN_{\alpha\alpha}(r,dr)$ denotes the number of $\alpha-\alpha$ pairs with radial distances falling within the interval $[r, r+dr]$. To obtain the temperature-dependent PDFs, the simulated supercell is first thermally equilibrated under the NPT-SCR ensemble for 500 ps at temperatures ranging from 2000 to 3200K in 200K increments. Subsequently, an additional 500-ps relaxation is performed under the NVT-BDP ensemble to further stabilize the system at the target temperatures. Following this, atomic trajectories are continuously sampled over a 1-ns production run under microcanonical (NVE) ensemble. The PDFs are statistically averaged over these trajectories up to a radial distance of 8.0 Å, employing a resolution of 500 bins to guarantee smooth curve profiles.

`

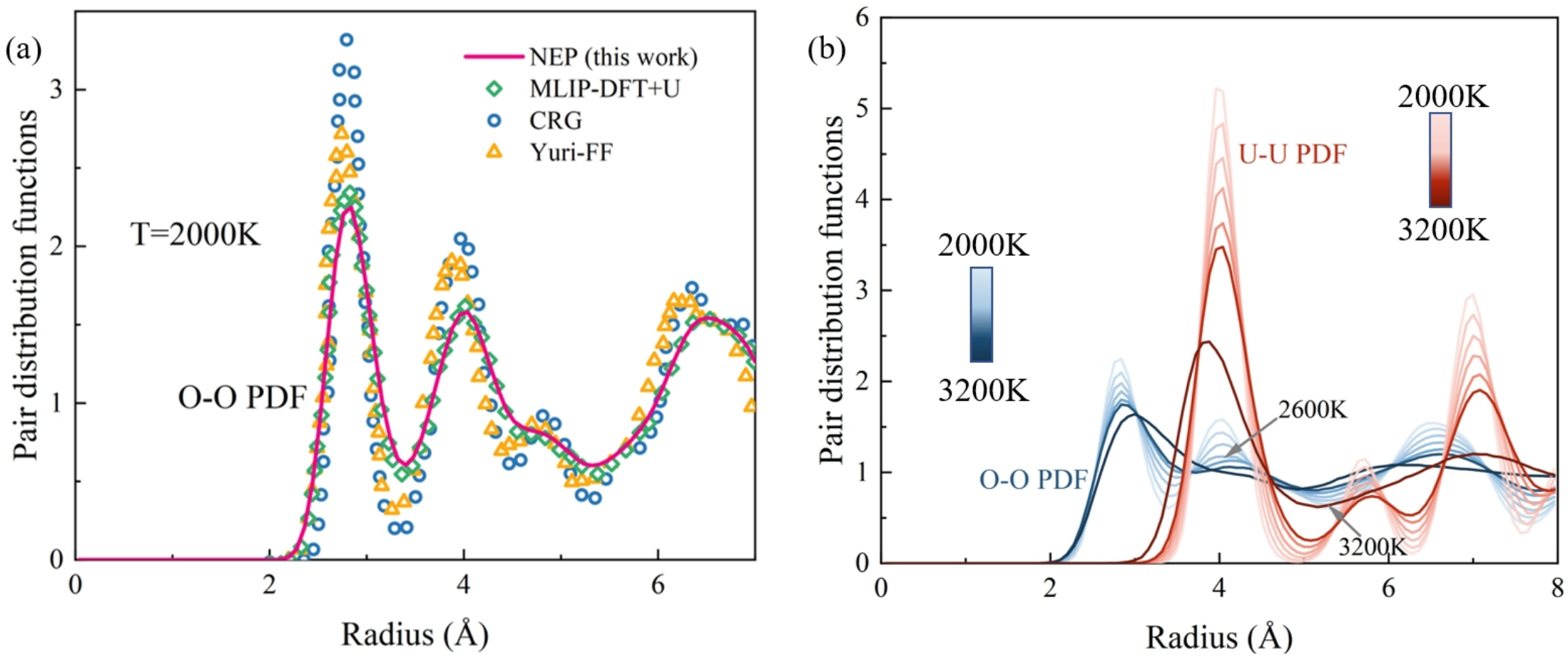

**Fig. 12**. Pair distribution functions (PDFs) of $UO_2$ as a function of radial distance. (a) The O-O PDF at 2000 K calculated by the NEP, compared against those by other interatomic potentials. (b) Temperature-dependent change of the O-O and U-U PDFs ranging from 2000 to 3200 K.

As provided in Fig. 12(a), the positions of the first and second peaks in the O-O PDF at 2000 K predicted by the NEP align with those derived from other interatomic potentials [10], validating the NEP's high fidelity in describing the local oxygen environment of $UO_2$. Notably, the slightly lower O-O PDF peak heights derived from our NEP indicate a more pronounced degree of thermally induced structural disorder. This behavior originates from the ST-targeted sampling employed during the training dataset construction, making the NEP more adept at capturing the disordering of oxygen sublattice. Fig. 12(b) illustrates the NEP-predicted temperature-dependent change of the U-U and O-O PDFs. As the temperature rises, the peak heights of both U-U and O-O pairs gradually decrease while their peak widths broaden. This phenomenon is primarily due to the greater structural disorder induced by enhanced thermal motion at elevated temperatures. Notably, the U-U PDF maintains clear first and second peaks up to 3000 K, with the second peak vanishing only at 3200K, demonstrating that the rigid fluorite cationic skeleton remains highly stable until the melting point ( $T_m$ = 3125 K). On the contrary, the O-O PDF undergoes a dramatic structural evolution at a lower temperature. Beyond 2600 K, the second peaks of the O-O pair flatten and merge, indicating a loss of long-range order as the oxygen sublattice transits into an amorphous,

`

liquid-like state. This observed transition temperature of 2600 K coincides with both the widely recognized empirical value of ~0.85 $T_m$ and the central temperature of the "λ-shape" LTEC peak (Fig. 10), firmly validating the NEP model's exceptional accuracy in capturing the superionic transition of $UO_2$.

**4.2 Kinetic decoupling of the cation and anion**

To further investigate the ions diffusion of $UO_2$, the mean squared displacement (MSD) of U-ion and O-ion are computed as,

$$\mathrm{MSD} = \left\langle \frac{1}{N}\sum_{i=1}^{N}(\mathbf{r}_i(t)-\mathbf{r}_i(0))^2 \right\rangle \tag{21}$$

where $N$ represents the total number of mobile ions, $\mathbf{r}_i(0)$ and $\mathbf{r}_i(t)$ are the position vectors of the *i*-th mobile particle at initial and current time, respectively. The MSDs of both the O and U ions are calculated under the NPT-SCR ensemble at temperatures ranging from 1600K to 3600K at 200K intervals. As presented in Fig. 13, the MSDs of both species increase as the temperature elevates. Clearly, the cationic MSD remains negligible until it rises sharply from zero at 3200 K (Fig. 13(b)), marking the onset of $UO_2$ melting. In contrast, the anionic MSD (Fig. 13(a)) undergoes two distinct abrupt increases above 2600 K and 3000 K (indicated by the arrows in Fig. 13), which correspond to the oxygen sublattice pre-melting and the subsequent $UO_2$ melting, respectively. Moreover, the MSDs of the O ions are substantially higher than those of the U ions at identical temperatures, indicative of the kinetic decoupling behavior of $UO_2$ arising from the large atomic mass difference between uranium and oxygen. This kinetic decoupling of anions and cations becomes particularly pronounced within the superionic state of $UO_2$, wherein O-ions migrate in a 'liquid-like' manner through the rigid 'skeleton' formed by U-ions. Such characteristic kinetic decoupling highlights the superionic state as a unique intermediate phase between the crystalline solid and fully molten liquid states [26, 27]. Crucially, these findings validate the NEP's capability to accurately capture the complex kinetic characteristics associated with ST onset in $UO_2$.

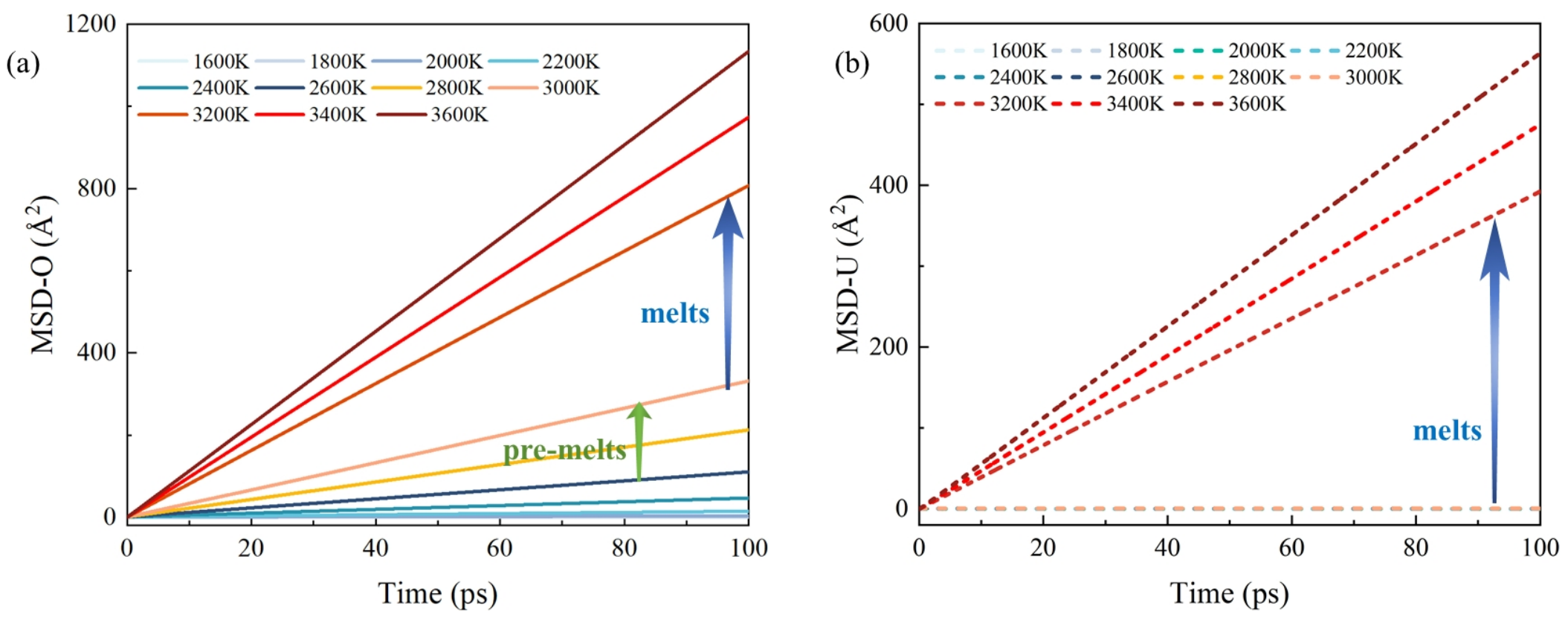

**Fig. 13.** Variations of the mean squared displacement (MSD) versus time for (a) O ions and, (b) U ions in $UO_2$ at temperatures ranging from 1600 to 3600K.The green and blue arrows signify the onset of oxygen sublattice pre-melting and the melting of $UO_2$, respectively.

### 4.3 Non-Arrhenius anionic diffusion

To provide a more quantitative analysis of the anionic diffusion behavior across various temperature regimes, the diffusion coefficient ( $D_O$ ) of O-ions is extracted from the corresponding MSD data via the Einstein relation as a function of production time $t$:

$$D_O = \mathrm{MSD} / 6t \tag{22}$$

Recognizing that the diffusion of O-ions operates as a thermally activated mechanism, its temperature-dependent evolution is well governed by the Arrhenius equation:

$$D_O = D_0 \exp(-Q_d / k_B T) \tag{23}$$

where $D_0$ is the pre-exponential factor, $k_B$ is the Boltzmann's constant, and $T$ is the absolute temperature. Therefore, the diffusion activation energy ( $Q_d$ ) is derived as,

$$Q_d = -\frac{\partial \ln D_O}{\partial (1 / k_B T)} \tag{24}$$

`

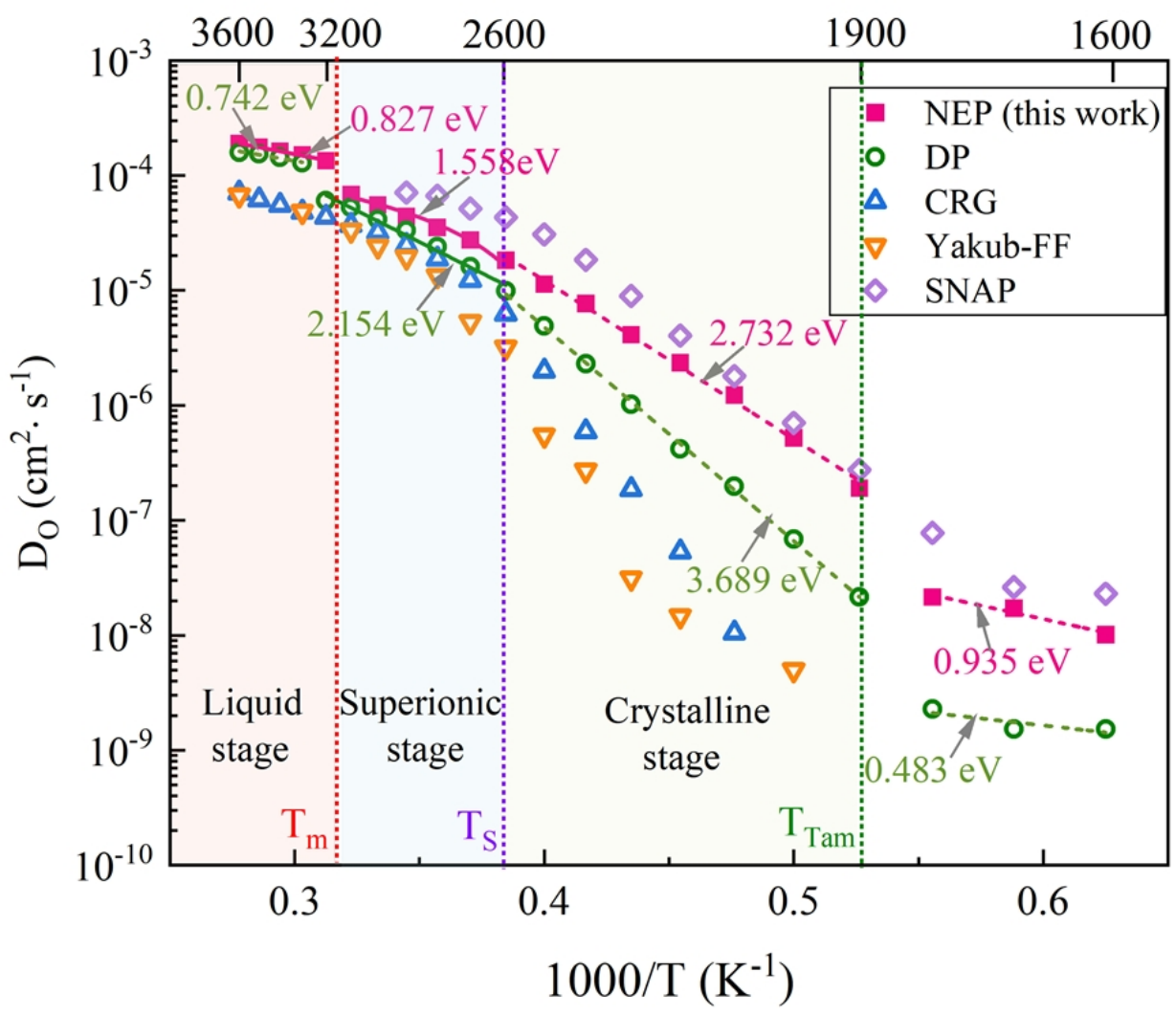

**Fig. 14**. Arrhenius plots of the O-ion diffusion coefficients ($D_O$) calculated by the NEP, compared with results from other interatomic potentials. The vertical dashed lines delineate three critical transition boundaries: the Tammann temperature ($T_{Tam}$ = 1900K), the superionic transition temperature ($T_S$ = 2600K), and the melting point ($T_m$ = 3200K). The entire temperature range is partitioned into distinct crystalline, superionic, and liquid regimes, each characterized by varying activation energies ($Q_d$)

Fig.14 presents the Arrhenius plots of the anionic diffusion coefficient ($D_O$) calculated using the NEP, in comparison with those obtained from other interatomic potentials. Evidently, the NEP-calculated $D_O$ values lie between those of SNAP (the highest values) and the other potentials (DP, CRG, and Yakub), indicating a reasonable anionic diffusivity reproduced by our NEP. Notably, the overall $\ln(D_O)$ versus $1000/T$ curve calculated by the NEP exhibits a non-linear characteristic across the entire temperature range, corresponding to a typical "non-Arrhenius" diffusion behavior. This "non-Arrhenius: trend is highly consistent with the other interatomic potentials displayed in Fig. 14, as well as with other published literatures [18, 19, 26]. Specifically, the overall Arrhenius plot can be partitioned into four distinct linear regimes with different slopes, segmented by three critical temperatures at 1900K, 2600K, and 3000K. These distinct slopes correspond to the respective diffusion activation energies ($Q_d$),

`

signifying successive transitions in the underlying diffusion mechanisms with rising temperature. The transition temperatures of 2600K and 3000K are readily interpretable, as they align precisely with the superionic transition temperature ($T_S$) and melting temperature ($T_m$) of $UO_2$, respectively, thereby delineating the boundaries for the superionic and liquid regimes. Notably, the transition temperature of 1900 K is close to the so-called Tammann temperature ($T_{Tam}$) of $UO_2$, a critical point where the solid lattice first exhibits significant ionic mobility [26]. Consequently, these three critical temperatures ($T_{Tam}$, $T_S$, and $T_m$) correspond to successive abrupt enhancements in anionic diffusivity. Furthermore, although the values of the activation energy ( $Q_d$ ) derived from our NEP are not identical to those of DP potential across different stages, their overall changing trends with temperature are highly consistent. Regarding the "non-Arrhenius" diffusion behavior of O ions, the low-temperature crystalline stage ($T<T_{Tam}$) exhibits a $Q_d$ of 0.935eV. This value falls within the documented experimental and theoretical calculations range of 0.5 to 3.3 eV, indicating that anionic diffusion is primarily governed by point-defect-facilitated jump processes [23]. As the temperature increases into the intermediate regime ($T_{Tam}<T<T_S$), a substantial number of oxygen Frenkel pairs (OFPs) begin to form. Such leads to a significant increase in the calculated $Q_d$ to 2.732 eV, reflecting the combined contributions of both defect formation and migration mechanisms. Upon entering the superionic stage ($T_S< T<T_m$), the $Q_d$ drops to 1.558 eV. This reduction stems from the saturation of OFP formation during the pre-melting of the oxygen sublattice, leaving solely the migration energy of OFP to dictate the $Q_d$ .[19, 106]. Finally, $Q_d$ experiences a further decline to 0.827eV within the liquid stage ($T>T_m$), corresponding to the highly mobile nature of the fully molten state. Overall, the successful reproduction of this "non-Arrhenius" diffusion behavior validates the reliability of our NEP in capturing the transitions of anionic diffusion mechanisms with temperatures increasing.

## 5. Conclusions:

`

(1) In this study, we propose a novel training workflow termed the "FAST" (**F**ine-tuning via **A**ctive-learning and **S**uperionic-**T**argeting) framework. This training strategy is designed to develop a highly robust and versatile neuroevolution potential (NEP) for $UO_2$, which remains applicable across a broad temperature range up to its melting point. Within the "FAST" framework, an initial baseline dataset comprising perturbed, sheared, defective, surface, and MD-sampled structures is prepared. Following this, we integrate the superionic transition (ST)-targeted sampling with successive active learning (AL)-enhanced exploration to construct a highly compact dataset containing merely 500 meticulously selected atomic configurations. Notably, during the high-accuracy DFT labelling of these configurations, we rigorously account for the strong correlation of uranium 5*f* electrons and the antiferromagnetic (AFM) ground state of $UO_2$. Utilizing this structurally diverse training dataset, a high-fidelity NEP for $UO_2$ is developed by fine-tuning the NEP89 foundation model for 20,000 steps. By systematic benchmarking the developed NEP against other interatomic potentials, first-principles calculations, and experimental measurements, we validate its exceptional predictive capability for several fundamental physical properties of $UO_2$.

(2) Specifically, the NEP accurately predicts several key static elastic and acoustic properties, including the elastic constants, bulk, shear and Young's modulus, as well as the longitudinal and shear sound velocities and the Debye temperature. The thermal softening and brittle-to-ductile transition (BDT) are also successfully captured through the temperature-dependent variations of the elastic constants and the yield-to-ultimate strength ratio ($\sigma_Y / \sigma_{UTS}$), respectively. Furthermore, our NEP is demonstrated to efficiently predict the energetics of point and extended defects in $UO_2$. The NEP-calculated values, which encompass the formation and migration energies of point defects alongside the generalized stacking fault energy surfaces, agree well with reported theoretical calculations and experimental data. The melting temperature of $UO_2$ predicted by the NEP is 3125 K, yielding a negligible deviation of only 0.16 % from the experimentally measured value of 3120 K. With this high-precision melting temperature serving as a prerequisite, the temperature-dependent lattice anharmonicity is fully investigated by calculating the linear thermal expansion coefficient (LTEC) and Grüneisen parameter at varying temperatures. The successful reproduction of the characteristic LTEC "λ-peak", signifies the exceptionally robust performance of our NEP in capturing the superionic (Bredig) transition at extremely elevated temperatures.

(3) More importantly, through comparative analyses of the pair distribution function (PDF) and the anionic diffusion coefficient ( $D_O$ ), the microscale origins of the ST are comprehensively elucidated. Mechanistically, at the onset of the superionic transition, the oxygen sublattice pre-melts while the uranium sublattice remains stable, leading to a kinetic decoupling between the U and O ions. This structural evolution and resulting kinetic discrepancy further trigger the occurrence of anomalous, "non-Arrhenius" anionic diffusion kinetics across the crystalline, superionic, and liquid states of $UO_2$.

In summary, this study develops a high-fidelity NEP capable of reproducing the complex thermo-mechanical-diffusion properties of $UO_2$ covering a wide temperature range. In addition, our proposed highly efficient and data-frugal "FAST" training framework may provide a highly transferable protocol and a valuable reference for developing robust machine learning interatomic potentials for other materials exhibiting similar complex phase transitions.

**Authorship contribution statement**

F.N. Zhuang: Investigation, Conceptualization, Validation, Writing – original draft, Writing – review & editing. G.S. Yan: Writing – review & editing, Supervision. H. Chen – Supervision & Funding acquisition. Y. Zhang – editing & review. W.S. Yu: Writing – original draft, Writing – review & editing, Supervision, Funding acquisition. M.L. Xu: Supervision & editing. S.P. Shen: Writing – review & editing, Supervision, Funding acquisition.

**Declaration of Competing Interest**

The authors declare that they have no known competing financial interests or personal relationships that could have appeared to influence the work reported in this paper.

**Acknowledgment**

The authors would like to express the gratitude for the support of NSFC (Grant No: 12202341 and 12527801). This study is supported by project of the Ministry of Industry and Information Technology.

**Reference:**

[1] F. Cappia, D. Pizzocri, M. Marchetti, A. Schubert, P. Van Uffelen, L. Luzzi, D. Papaioannou, R. Macián-Juan, V. Rondinella, Microhardness and Young's modulus of high burn-up $UO_2$ fuel, J. Nucl. Mater. 479 (2016) 447-454.

[2] R. Henry, I. Zacharie-Aubrun, T. Blay, N. Tarisien, S. Chalal, X. Iltis, J.-M. Gatt, C. Langlois, S. Meille, Irradiation effects on the fracture properties of $UO_2$ fuels studied by micro-mechanical testing, J. Nucl. Mater. 536 (2020) 152179.

[3] D. Horlait, J. Domange, M.-L. Amany, M. Gérardin, M.-F. Barthe, G. Carlot, E. Gilabert, Experimental investigation of Kr diffusion in $UO_{2+x}$: slight deviations from stoichiometry, significant effects on diffusion kinetics and mechanisms, J. Nucl. Mater. 574 (2023) 154191.

[4] M. Saidy, W.H. Hocking, J.F. Mouris, P. Garcia, G. Carlot, B. Pasquet, Thermal diffusion of iodine in $UO_2$ and $UO_{2+x}$, J. Nucl. Mater. 372(2-3) (2008) 405-415.

[5] J. Harding, D. Martin, A recommendation for the thermal conductivity of $UO_2$, J. Nucl. Mater. 166(3) (1989) 223-226.

[6] P. Lucuta, I. Hastings, A pragmatic approach to modelling thermal conductivity of irradiated $UO_2$ fuel: review and recommendations, J. Nucl. Mater. 232(2-3) (1996) 166-180.

[7] B. Dorado, B. Amadon, M. Freyss, M. Bertolus, DFT+ U calculations of the ground state and metastable states of uranium dioxide, Phys. Rev. B 79(23) (2009) 235125.

[8] E. Vathonne, J. Wiktor, M. Freyss, G. Jomard, M. Bertolus, DFT+ U investigation of charged point defects and clusters in $UO_2$, J. Phys. Condens. Matter 26(32) (2014) 325501.

[9] B. Dorado, D.A. Andersson, C.R. Stanek, M. Bertolus, B.P. Uberuaga, G. Martin, M. Freyss, P. Garcia, First-principles calculations of uranium diffusion in uranium dioxide, Phys. Rev. B 86(3) (2012) 035110.

[10] E. Stippell, L. Alzate-Vargas, K.N. Subedi, R.M. Tutchton, M.W. Cooper, S. Tretiak, T. Gibson, R.A. Messerly, Building a DFT+ U machine learning interatomic potential for uranium dioxide,

`

Artif. Intell. Chem. 2(1) (2024) 100042.
[11] C. Basak, A. Sengupta, H. Kamath, Classical molecular dynamics simulation of $UO_2$ to predict thermophysical properties, J. Alloy. Compd. 360(1-2) (2003) 210-216.
[12] N.-D. Morelon, D. Ghaleb, J.-M. Delaye, L. Van Brutzel, A new empirical potential for simulating the formation of defects and their mobility in uranium dioxide, Philos. Mag. 83(13) (2003) 1533-1555.
[13] E. Yakub, C. Ronchi, D. Staicu, Molecular dynamics simulation of premelting and melting phase transitions in stoichiometric uranium dioxide, J. Chem. Phys. 127(9) (2007).
[14] M. Cooper, M. Rushton, R. Grimes, A many-body potential approach to modelling the thermomechanical properties of actinide oxides, J. Phys. Condens. Matter 26(10) (2014) 105401.
[15] A. Soulié, J.-P. Crocombette, A. Kraych, F. Garrido, G. Sattonnay, E. Clouet, Atomistically-informed thermal glide model for edge dislocations in uranium dioxide, Acta Mater. 150 (2018) 248-261.
[16] X.-Y. Liu, M.W.D. Cooper, K.J. McClellan, J.C. Lashley, D.D. Byler, B. Bell, R. Grimes, C.R. Stanek, D.A. Andersson, Molecular dynamics simulation of thermal transport in $UO_2$ containing uranium, oxygen, and fission-product defects, Phys. Rev. Appl. 6(4) (2016) 044015.
[17] H. Balboa, L. Van Brutzel, A. Chartier, Y. Le Bouar, Assessment of empirical potential for MOX nuclear fuels and thermomechanical properties, J. Nucl. Mater. 495 (2017) 67-77.
[18] F. Zhuang, G. Yan, H. Chen, Y. Zhang, W. Yu, M. Xu, S. Shen, Superionic Transition-Induced Diffusional Creep Behavior Transitions in Nanocrystalline $(U_{1-x}, Th_x)O_2$ Nuclear Fuels, Acta Mech. Solida Sin. (2026) 1-15.
[19] F. Zhuang, G. Yan, W. Yu, M. Xu, S. Shen, The high-temperature creep behavior transition in stoichiometric nanocrystalline $(Pu_x, U_{1-x})O_2$ mixed oxides, J. Am. Ceram. Soc. 108(4) (2025) e20275.
[20] C. Chil, A. Pivano, B. Michel, E. Bourasseau, Fracture behavior of grain boundaries in (U, Pu) $O_2$ fuels: An atomistic study, J. Am. Ceram. Soc. 109(1) (2026) e70326.
[21] J. Behler, Perspective: Machine learning potentials for atomistic simulations, J. Chem. Phys. 145(17) (2016).
[22] E.T. Dubois, J. Tranchida, J. Bouchet, J.-B. Maillet, Atomistic simulations of nuclear fuel $UO_2$ with machine learning interatomic potentials, Phys. Rev. Mater. 8(2) (2024) 025402.
[23] K. Konashi, N. Kato, K. Mori, K. Kurosaki, Neural network potential for molecular dynamics calculation of UO2, J. Nucl. Mater. 607 (2025) 155660.
[24] J. Zhong, L. Zhang, T. Bo, Active learning-enhanced neuroevolution potential for predictive modeling of $UO_2$ thermophysical properties, Phys. Chem. Chem. Phys. 28(1) (2026) 671-682.
[25] D.H. Hurley, A. El-Azab, M.S. Bryan, M.W. Cooper, C.A. Dennett, K. Gofryk, L. He, M. Khafizov, G.H. Lander, M.E. Manley, Thermal energy transport in oxide nuclear fuel, Chem. Rev. 122(3) (2022) 3711-3762.
[26] H. Zhang, X. Wang, A. Chremos, J.F. Douglas, Superionic $UO_2$: A model anharmonic crystalline material, J. Chem. Phys. 150(17) (2019).
[27] K. Naito, High-temperature heat capacities of $UO_2$ and doped $UO_2$, J. Nucl. Mater. 167 (1989) 30-35.
[28] J. Hiernaut, G. Hyland, C. Ronchi, Premelting transition in uranium dioxide, Int. J. Thermophys. 14(2) (1993) 259-283.
[29] A.V. Lunev, B.A. Tarasov, A classical molecular dynamics study of the correlation between the

Bredig transition and thermal conductivity of stoichiometric uranium dioxide, J. Nucl. Mater. 415(2) (2011) 217-221.
[30] P.C. Fossati, P.A. Burr, M.W. Cooper, C.O. Galvin, R.W. Grimes, Superionic transition in uranium dioxide: Insights from molecular dynamics and lattice dynamics simulations, Phys. Rev. Mater. 8(11) (2024) 115404.
[31] P. Ying, C. Qian, R. Zhao, Y. Wang, K. Xu, F. Ding, S. Chen, Z. Fan, Advances in modeling complex materials: The rise of neuroevolution potentials, Chem. Phys. Rev. 6(1) (2025).
[32] X. Liu, K. Zeng, Z. Luo, Y. Wang, T. Zhao, Z. Xu, Fine-tuning universal machine-learned interatomic potentials: A tutorial on methods and applications, J. Appl. Phys. 139(4) (2026).
[33] T. Liang, K. Xu, E. Lindgren, Z. Chen, R. Zhao, J. Liu, E. Berger, B. Tang, B. Zhang, Y. Wang, NEP89: Universal neuroevolution potential for inorganic and organic materials across 89 elements, arXiv preprint arXiv:2504.21286 (2025).
[34] Z. Fan, Y. Wang, P. Ying, K. Song, J. Wang, Y. Wang, Z. Zeng, K. Xu, E. Lindgren, J.M. Rahm, GPUMD: A package for constructing accurate machine-learned potentials and performing highly efficient atomistic simulations, J. Chem. Phys. 157(11) (2022).
[35] C.O. Galvin, M. Machida, H. Nakamura, D.A. Andersson, M.W. Cooper, Correlations for the specific heat capacity of $(U_xPu_{1-x})_{1-\gamma}Gd_\gamma O_{2-}z$ derived from molecular dynamics, J. Nucl. Mater. 572 (2022) 154028.
[36] S.V. Ushakov, A. Navrotsky, R.J. Weber, J.C. Neuefeind, Structure and thermal expansion of YSZ and $La_2Zr_2O_7$ above 1500 °C from neutron diffraction on levitated samples, J. Am. Ceram. Soc. 98(10) (2015) 3381-3388.
[37] Y.-C. Yin, J.-T. Yang, J.-D. Luo, G.-X. Lu, Z. Huang, J.-P. Wang, P. Li, F. Li, Y.-C. Wu, T. Tian, A $LaCl_3$-based lithium superionic conductor compatible with lithium metal, Nature 616(7955) (2023) 77-83.
[38] K. Jun, Y. Chen, G. Wei, X. Yang, G. Ceder, Diffusion mechanisms of fast lithium-ion conductors, Nat. Rev. Mater. 9(12) (2024) 887-905.
[39] Y. Kato, S. Hori, T. Saito, K. Suzuki, M. Hirayama, A. Mitsui, M. Yonemura, H. Iba, R. Kanno, High-power all-solid-state batteries using sulfide superionic conductors, Nat. Energy 1(4) (2016) 16030.
[40] W. Liu, Y. Zhou, Energy transport in superionic crystals, Phys. Rev. Lett. 134(14) (2025) 146301.
[41] B. Li, H. Wang, Y. Kawakita, Q. Zhang, M. Feygenson, H. Yu, D. Wu, K. Ohara, T. Kikuchi, K. Shibata, Liquid-like thermal conduction in intercalated layered crystalline solids, Nat. Mater. 17(3) (2018) 226-230.
[42] M. Millot, F. Coppari, J.R. Rygg, A. Correa Barrios, S. Hamel, D.C. Swift, J.H. Eggert, Nanosecond X-ray diffraction of shock-compressed superionic water ice, Nature 569(7755) (2019) 251-255.
[43] Y. He, W. Zhang, Q. Hu, S. Sun, J. Hu, D. Liu, L. Zhou, L. Dai, D.Y. Kim, S.A. Redfern, Absence of dehydration due to superionic transition at Earth's core-mantle boundary, Sci. Adv. 12(5) (2026) eaeb3006.
[44] C. Chen, Y. Li, R. Zhao, Z. Liu, Z. Fan, G. Tang, Z. Wang, NepTrain and NepTrainKit: Automated active learning and visualization toolkit for neuroevolution potentials, Comput. Phys. Commun. (2025) 109859.
[45] A.P. Thompson, H.M. Aktulga, R. Berger, D.S. Bolintineanu, W.M. Brown, P.S. Crozier, P.J.

In't Veld, A. Kohlmeyer, S.G. Moore, T.D. Nguyen, LAMMPS-a flexible simulation tool for particle-based materials modeling at the atomic, meso, and continuum scales, Comput. Phys. Commun. 271 (2022) 108171.
[46] K. Xu, H. Bu, S. Pan, E. Lindgren, Y. Wu, Y. Wang, J. Liu, K. Song, B. Xu, Y. Li, GPUMD 4.0: a high-performance molecular dynamics package for versatile materials simulations with machine-learned potentials, Mater. Genome Eng. Adv. 3(3) (2025) e70028.
[47] M. Cooper, S. Murphy, M. Rushton, R. Grimes, Thermophysical properties and oxygen transport in the ($U_x$, $Pu_{1-x}$)$O_2$ lattice, J. Nucl. Mater. 461 (2015) 206-214.
[48] Z. Yan, D. Li, X. Wu, Z. Liu, C. Hua, B. Situ, H. Yang, S. Tang, B. Tang, Z. Wang, GPUMDkit: A User-Friendly Toolkit for GPUMD and NEP, Mater. Genome Eng. Adv. (2026) e70074.
[49] Y. Lysogorskiy, A. Bochkarev, M. Mrovec, R. Drautz, Active learning strategies for atomic cluster expansion models, Phys. Rev. Mater. 7(4) (2023) 043801.
[50] L. Zhang, D.-Y. Lin, H. Wang, R. Car, W. E, Active learning of uniformly accurate interatomic potentials for materials simulation, Phys. Rev. Mater. 3(2) (2019) 023804.
[51] G. Kresse, J. Furthmüller, Efficient iterative schemes for ab initio total-energy calculations using a plane-wave basis set, Phys. Rev. B 54(16) (1996) 11169.
[52] I.C. Njifon, E. Torres, Atomistic investigation of uranium oxycarbide (UCO) phase stability at finite temperatures: DFT+ U and AIMD+ U approaches, J. Nucl. Mater. 554 (2021) 153046.
[53] G. Kresse, D. Joubert, From ultrasoft pseudopotentials to the projector augmented-wave method, Phys. Rev. B 59(3) (1999) 1758.
[54] J.P. Perdew, K. Burke, M. Ernzerhof, Generalized gradient approximation made simple, Phys. Rev. Lett. 77(18) (1996) 3865.
[55] H.J. Monkhorst, J.D. Pack, Special points for Brillouin-zone integrations, Phys. Rev. B 13(12) (1976) 5188.
[56] S. Dudarev, D.N. Manh, A. Sutton, Effect of Mott-Hubbard correlations on the electronic structure and structural stability of uranium dioxide, Philos. Mag. B 75(5) (1997) 613-628.
[57] J.P. Allen, G.W. Watson, Occupation matrix control of d-and f-electron localisations using DFT+ U, Phys. Chem. Chem. Phys. 16(39) (2014) 21016-21031.
[58] A. Claisse, M. Klipfel, N. Lindbom, M. Freyss, P. Olsson, GGA+ U study of uranium mononitride: A comparison of the U-ramping and occupation matrix schemes and incorporation energies of fission products, J. Nucl. Mater. 478 (2016) 119-124.
[59] J.F. Ziegler, J.P. Biersack, The stopping and range of ions in matter, Treat. Heavy-Ion Sci. 6, Springer, 1985, pp. 93-129.
[60] J. Liu, J. Byggmästar, Z. Fan, P. Qian, Y. Su, Large-scale machine-learning molecular dynamics simulation of primary radiation damage in tungsten, Phys. Rev. B 108(5) (2023) 054312.
[61] G. Bussi, D. Donadio, M. Parrinello, Canonical sampling through velocity rescaling, J. Chem. Phys. 126(1) (2007).
[62] M. Bernetti, G. Bussi, Pressure control using stochastic cell rescaling, J. Chem. Phys. 153(11) (2020).
[63] A. Hjorth Larsen, J. Jørgen Mortensen, J. Blomqvist, I.E. Castelli, R. Christensen, M. Dułak, J. Friis, M.N. Groves, B. Hammer, C. Hargus, The atomic simulation environment—a Python library for working with atoms, J. Phys. Condens. Matter 29(27) (2017) 273002.
[64] E. Lindgren, M. Rahm, E. Fransson, F. Eriksson, N. Österbacka, Z. Fan, P. Erhart, calorine: A Python package for constructing and sampling neuroevolution potential models, J. Open Source

Softw. 9(95) (2024) 6264.
[65] J. Fink, Thermophysical properties of uranium dioxide, J. Nucl. Mater. 279(1) (2000) 1-18.
[66] E. Brando, E. Dubois, M. Rochedy, J. Bouchet, F. Bruneval, J. Crocombette, J. Tranchida, Atomistic investigation of the temperature and stoichiometry dependence of the heat-capacity in nuclear fuel $UO_{2+x}$, J. Nucl. Mater. (2025) 156280.
[67] J.-L. Chen, N. Kaltsoyannis, DFT+ U study of uranium dioxide and plutonium dioxide with occupation matrix control, J. Phys. Chem. C 126(27) (2022) 11426-11435.
[68] Z.-G. Mei, M. Stan, J. Yang, First-principles study of thermophysical properties of uranium dioxide, J. Alloy. Compd. 603 (2014) 282-286.
[69] J. Wachtman Jr, M. Wheat, H. Anderson, J. Bates, Elastic constants of single crystal $UO_2$ at 25 °C, J. Nucl. Mater. 16(1) (1965) 39-41.
[70] G. Leinders, T. Cardinaels, K. Binnemans, M. Verwerft, Accurate lattice parameter measurements of stoichiometric uranium dioxide, J. Nucl. Mater. 459 (2015) 135-142.
[71] I. Fritz, Elastic properties of $UO_2$ at high pressure, J. Appl. Phys. 47(10) (1976) 4353-4358.
[72] M. Parrinello, A. Rahman, Strain fluctuations and elastic constants, J. Chem. Phys. 76(5) (1982) 2662-2666.
[73] M.T. Hutchings, High-temperature studies of $UO_2$ and $ThO_2$ using neutron scattering techniques, J. Chem. Soc. Faraday Trans. 2 83(7) (1987) 1083-1103.
[74] G.J. Martyna, D.J. Tobias, M.L. Klein, Constant pressure molecular dynamics algorithms, J. Chem. Phys. 101(4177) (1994) 10.1063.
[75] Y. Zhang, X.-Y. Liu, P.C. Millett, M. Tonks, D.A. Andersson, B. Biner, Crack tip plasticity in single crystal $UO_2$: Atomistic simulations, J. Nucl. Mater. 430(1-3) (2012) 96-105.
[76] P.C. Fossati, L. Van Brutzel, A. Chartier, J.-P. Crocombette, Simulation of uranium dioxide polymorphs and their phase transitions, Phys. Rev. B 88(21) (2013) 214112.
[77] E. Rapperport, A. Huntress, Deformation modes of single crystal uranium dioxide from 700 °C to 1900 °C, United States Atomic Energy Commission, Office of Technical Information, 1960.
[78] R. Robins, P. Baldock, Uranium oxide cleavage, J. Am. Ceram. Soc. 43 (1960).
[79] T. Arima, S. Yamasaki, Y. Inagaki, K. Idemitsu, Evaluation of thermal properties of $UO_2$ and $PuO_2$ by equilibrium molecular dynamics simulations from 300 to 2000 K, J. Alloy. Compd. 400(1-2) (2005) 43-50.
[80] S. Potashnikov, A. Boyarchenkov, K. Nekrasov, A.Y. Kupryazhkin, High-precision molecular dynamics simulation of $UO_2$–$PuO_2$: Pair potentials comparison in $UO_2$, J. Nucl. Mater. 419(1-3) (2011) 217-225.
[81] J. Roberts, B. Wrona, Nature of brittle-to-ductile transition in $UO_2$-20 wt% $PuO_2$ nuclear fuel, J. Nucl. Mater. 41(1) (1971) 23-38.
[82] B. Dorado, J. Durinck, P. Garcia, M. Freyss, M. Bertolus, An atomistic approach to self-diffusion in uranium dioxide, J. Nucl. Mater. 400(2) (2010) 103-106.
[83] D. Bathellier, L. Messina, M. Freyss, M. Bertolus, T. Schuler, M. Nastar, P. Olsson, E. Bourasseau, Effect of cationic chemical disorder on defect formation energies in uranium – plutonium mixed oxides, J. Appl. Phys. 132(17) (2022).
[84] H. Matzke, Atomic transport properties in $UO_2$ and mixed oxides $(U, Pu)O_2$, J. Chem. Soc. Faraday Trans. 2 83(7) (1987) 1121-1142.
[85] X.-Y. Liu, D. Andersson, Small uranium and oxygen interstitial clusters in $UO_2$: An empirical potential study, J. Nucl. Mater. 547 (2021) 152783.

[86] S.M. Zamzamian, A. Zolfaghari, Z. Kowsar, Molecular dynamics investigation of xenon, uranium, and oxygen diffusion in $UO_2$ nuclear fuel, Comput. Mater. Sci. 211 (2022) 111553.
[87] T. Arima, K. Yoshida, K. Idemitsu, Y. Inagaki, I. Sato, Molecular dynamics analysis of diffusion of uranium and oxygen ions in uranium dioxide, IOP Conf. Ser. Mater. Sci. Eng. 2010, p. 012003.
[88] X.-M. Bai, A. El-Azab, J. Yu, T.R. Allen, Migration mechanisms of oxygen interstitial clusters in $UO_2$, J. Phys. Condens. Matter 25(1) (2013) 015003.
[89] B. Dorado, P. Garcia, G. Carlot, C. Davoisne, M. Fraczkiewicz, B. Pasquet, M. Freyss, C. Valot, G. Baldinozzi, D. Siméone, First-principles calculation and experimental study of oxygen diffusion in uranium dioxide, Phys. Rev. B 83(3) (2011) 035126.
[90] F. Gupta, A. Pasturel, G. Brillant, Diffusion of oxygen in uranium dioxide: A first-principles investigation, Phys. Rev. B 81(1) (2010) 014110.
[91] Y. Yun, P.M. Oppeneer, H. Kim, K. Park, Defect energetics and Xe diffusion in $UO_2$ and $ThO_2$, Acta Mater. 57(5) (2009) 1655-1659.
[92] R. Jackson, A. Murray, J. Harding, C. Catlow, The calculation of defect parameters in $UO_2$, Philos. Mag. A 53(1) (1986) 27-50.
[93] C.R.A. Catlow, Point defect and electronic properties of uranium dioxide, Proc. R. Soc. Lond. A 353(1675) (1977) 533-561.
[94] J. Belle, Oxygen and uranium diffusion in uranium dioxide (a review), J. Nucl. Mater. 30(1-2) (1969) 3-15.
[95] K. Ashbee, C. Yust, A mechanism for the ease of slip in $UO_{2+x}$, J. Nucl. Mater. 110(2-3) (1982) 246-250.
[96] L.S. Liyanage, S.-G. Kim, J. Houze, S. Kim, M.A. Tschopp, M.I. Baskes, M.F. Horstemeyer, Structural, elastic, and thermal properties of cementite ($Fe_3C$) calculated using a modified embedded atom method, Phys. Rev. B 89(9) (2014) 094102.
[97] C. Guéneau, A. Chartier, P. Fossati, L. Van Brutzel, P. Martin, Thermodynamic and thermophysical properties of the actinide oxides, (2020).
[98] J.W. Pang, A. Chernatynskiy, B.C. Larson, W.J. Buyers, D.L. Abernathy, K.J. McClellan, S.R. Phillpot, Phonon density of states and anharmonicity of $UO_2$, Phys. Rev. B 89(11) (2014) 115132.
[99] A. Dworkin, M.A. Bredig, Diffuse transition and melting in fluorite and antifluorite type of compounds. Heat content of potassium sulfide from 298 to 1260. degree. K, J. Phys. Chem. 72(4) (1968) 1277-1281.
[100] T.R. Pavlov, M. Wenman, L. Vlahovic, D. Robba, R. Konings, P. Van Uffelen, R. Grimes, Measurement and interpretation of the thermo-physical properties of $UO_2$ at high temperatures: The viral effect of oxygen defects, Acta Mater. 139 (2017) 138-154.
[101] D. Martin, The thermal expansion of solid $UO_2$ and (U, Pu) mixed oxides—a review and recommendations, J. Nucl. Mater. 152(2-3) (1988) 94-101.
[102] I.C. Njifon, E. Torres, Phonons and thermophysical properties of $U_{1-\gamma}Pu_{\gamma}O_2$ mixed oxide (MOX) fuels, J. Nucl. Mater. 537 (2020) 152158.
[103] T. Yamashita, N. Nitani, T. Tsuji, T. Kato, Thermal expansion of neptunium-uranium mixed oxides, J. Nucl. Mater. 247 (1997) 90-93.
[104] H. Serizawa, Y. Arai, Y. Suzuki, Simultaneous determination of X-ray Debye temperature and Grüneisen constant for actinide dioxides: $PuO_2$ and $ThO_2$, J. Nucl. Mater. 280(1) (2000) 99-105.
[105] B. Willis, Neutron diffraction studies of the actinide oxides II. Thermal motions of the atoms in uranium dioxide and thorium dioxide between room temperature and 1100 °C, Proc. R. Soc.

Lond. A 274(1356) (1963) 134-144.
[106] M.W. Cooper, S.T. Murphy, P. Fossati, M.J. Rushton, R.W. Grimes, Thermophysical and anion diffusion properties of $(U_x, Th_{1-x})O_2$, Proc. R. Soc. A 470(2171) (2014).

`